\documentclass[pra,showpacs,twocolumn,amsmath,amssymb,floatfix]{revtex4}
\usepackage{amsthm,times,mathptmx}
\usepackage{latexsym}
\usepackage{amsfonts}
\usepackage{bbm,dsfont}
\usepackage{graphicx,color}
\newcommand{\R}{\mathbb{R}}
\newcommand{\C}{\mathbb{C}}

\newcommand{\Id}{\mathbb{I}}
\newcommand{\hil}{{\cal H}}
\newcommand{\tr}[1]{\mathrm{Tr}\left[ {#1} \right]} 
 
\newcommand{\ket}[1]{\left\vert {#1} \right\rangle}

\newcommand{\ip}{{\scriptscriptstyle IN}}
\newcommand{\ot}{{\scriptscriptstyle OUT}}
\def\OPA{{\rm OPA}}
\def\BS{{\rm BS}}
\def\Nin{N_{\rm in}}
\def\Xab{X_{ab}}

\begin{document}
\title{Gaussian state interferometry with passive and active elements}
\author{Carlo Sparaciari}\email{carlo.sparaciari.14@ucl.ac.uk}
\affiliation{Department of Physics \& Astronomy, 
University College London, London, WC1E 6BT, United Kingdom.}
\author{Stefano Olivares}\email{stefano.olivares@mi.infn.it}
\affiliation{Dipartimento di Fisica dell'Universit\`a degli Studi di Milano,
I-20133 Milano, Italia}
\affiliation{INFN Sezione di Milano, I-20133 Milano, Italy}
\author{Matteo G. A. Paris}\email{matteo.paris@fisica.unimi.it}
\affiliation{Dipartimento di Fisica dell'Universit\`a degli Studi di Milano,
I-20133 Milano, Italia}
\affiliation{INFN Sezione di Milano, I-20133 Milano, Italy}
\begin{abstract}
We address precision of optical interferometers 
fed by Gaussian states and involving passive and/or active elements, 
such as beam splitters, photodetectors and optical parametric amplifiers.  
We first address the ultimate bounds to precision by discussing 
the behaviour of the quantum Fisher information.
We then consider photodetection at the output and calculate 
the sensitivity of the interferometers taking into account 
the non unit quantum efficiency of the detectors.   
Our results show that in the ideal case of photon number
detectors with unit quantum efficiency the best configuration is the
symmetric one, namely, passive (active) interferometer with passive
(active) detection stage: in this case one may achieve Heisenberg
scaling of sensitivity by suitably optimizing over Gaussian states at
the input. On the other hand, in the realistic case of detectors with 
non unit quantum efficiency, the performances of passive scheme are 
unavoidably degraded, whereas detectors involving optical parametric 
amplifiers allow to fully compensate the presence of loss in the 
detection stage, thus restoring the Heisenberg scaling.
\end{abstract}
\date{\today}
\maketitle
\section{Introduction}\label{s1:int}
Optical interferometry is a mature quantum technology
\cite{raf:rev,rst02}. 
Results in this field nicely shows how quantum features of 
information carriers may improve performances of devices previously 
based on classical signals. In the last decades, many efforts have 
been made in order to find the ultimate limits to interferometric 
precision, but only recently quantum enhancement of sensitivity 
using squeezed light has been demonstrated \cite{grav:11,raf14,feedphase}.
As a matter of fact, the presence of losses such as a non unit quantum
efficiency in the detection stage, limits the performances of 
interferometers. The interferometric sensitivity, which ideally
achieve Heisenberg scaling upon exploiting squeezing, 
may be degraded in the presence of losses, even down to the shot 
noise limit \cite{hil93,dar94,par95,kim99,oli:par:OptSp,pez08}.  
\par 
Much attention has been devoted so far to  
Mach-Zehnder-like interferometers based on 
passive devices, such as beam splitters, in which squeezed photons 
are injected as input states. On the other hand, promising 
results have been obtained exploiting active elements, such as 
optical parametric amplifiers. The so-called SU$(1,1)$ 
interferometers \cite{yur86,mar:12,bar14} and the coherent-light-boosted 
interferometers \cite{kim09,agar:10} belong to this class. 
Quantum enhaced precision in active
interferometers have been recently demostrated \cite{wp12,act14}.
\par
In this paper we build on the results obtained in Ref.~\cite{CS:JOSAB}
where the ultimate limits to interferometric precision have been
assessed by the tools of quantum estimation theory \cite{par09}.  Both
passive and active interferometers fed by Gaussian states have been
considered, and the corresponding bound to precision have been obtained
by maximizing the quantum Fisher information over the possible input
signals. Results suggest that Heisenberg scaling with optimized
constant may be achieved with suitably adjusted Gaussian signals. 
On the other  hand, the optimal measurements suggested by quantum
estimation theory may not be feasible with current technology and
thus it becomes relevant to assess the performances of active and
passive interferometers when a specific, and realistic, detection stage is
considered. We focus on photon number detection, assisted either by
passive elements or active ones, also taking into account the effects of
imperfections, i.e. non unit quantum efficiency, in the detection
process. Indeed, imperfections at the detection stage represent the 
major limits to interferometric precision. In addition, since we are 
going to consider Gaussian signals and devices, other losses within
the interferometer may be subsumed by an overall quantum efficiency
\cite{pau93}.
\par
In a passive detection scheme, the two beams outgoing the
interferometers are mixed at a beam splitter before the detection, in
the active counterpart, the beams interact through an optical parametric
amplifiers and are finally detected. We carry out the optimization over
the input states and, in turn, found the optimal working regimes in 
both the ideal case and for non unit quantum efficiency.
As we will see, in the ideal case of photon number
detectors with unit quantum efficiency Heisenberg scaling is achieved 
using symmetric configurations, i.e. by passive interferometers 
with passive detection stages or by the active/active counterpart. 
On the other hand, in the realistic case of detectors with 
non unit quantum efficiency, the performances of passive schemes are 
unavoidably degraded, whereas active detectors involving parametric 
amplifiers allow to compensate for presence of loss in the detection 
stage, as it happens in single-photon active interferometry 
\cite{scia10}, and restoring the Heisenberg scaling, 
\par
The structure of the paper is as follows.  In Sect.~\ref{s2:lqet}, we
briefly summarize the main tools of quantum estimation theory, whereas
in Sect.~\ref{s3:api} we discuss the main features of passive and
active interferometers.  The ultimate limits to the interferometry
precision for both active and passive schemes are described in
Sect.~\ref{s4:qfii}.  In Sect.~\ref{s5:sapi} we evaluate and optimize
the sensitivity for the passive and active interferometers when a
passive or an active measurement stage is considered. In particular, we
show that an active device can compensate the losses due to a non-unit
quantum efficiency restoring the ideal case sensitivity achieved with
lossless detectors.  This features is discussed in details in
Sect.~\ref{s6:feat}, whereas sect.~\ref{s7:con} closes the paper with
some concluding remarks.
\section{Local Quantum Estimation Theory}\label{s2:lqet}
Let's consider a parameter $\phi$ which can not be directly measured,
i.e. it does not corresponds to a quantum observable, and a quantum 
system described by the density operator $\rho_{\phi} \in 
\mathcal{S}(\hil)$ carrying
information about it, where $\hil$ is the Hilbert space associated with
the system. An inference strategy, and in turn an estimate for $\phi$, 
may be obtained through repeated measurements of a quantum
observable on $\rho_{\phi}$ followed by a suitable classical data 
processing on the measurement results. We describe the observable with 
a positive operator-valued measure (POVM)  
$E : \mathcal{B}(\Lambda) \rightarrow \mathcal{B}(\hil)$, where 
$\Lambda$ is the set of all the possible measurement outcomes,
$\mathcal{B}(\Lambda)$ is the Borel $\sigma$-algebra on $\Lambda$ and
$\mathcal{B}(\hil)$ is the set of bounded operator in $\hil$.
\par
The function of data providing the value of $\phi$ is usually referred
to as an {\em estimator} and its variance over data represents the 
uncertainty $\Delta^2 \phi$ of the overall estimation procedure, 
which in turn determines its precision. In particular, we focus on the 
lower bound of the
achievable precision in the estimation of $\phi$.  This bound is
provided by the Cram\'er-Rao theorem, which states that
\begin{equation}\label{CRb}
\langle \Delta^2 \phi \rangle \geq \frac{1}{M F(\phi)}
\end{equation}
where $\langle \Delta^2 \phi \rangle$ is the variance of $\phi$,
$\langle \cdots \rangle = {\rm Tr}[\rho_\phi \cdots ]$, $M$ is the
number of repeated measurements performed on the system, and $F(\phi)$
is the {\it Fisher Information} (FI):
\begin{equation}\label{FI}
F(\phi) = \int_{\Lambda} \mathrm{d}x\, p(x \vert \phi) \left[
\partial_\phi \log p(x \vert \phi) \right]^2 \end{equation}
where $p(x \vert \phi) = \tr{\rho_{\phi}\, E(x)}$ is the probability
distribution of the outcome $x$ conditional on the unknown actual value
$\phi$ of the parameter and $E(x)$ is the element of the POVM 
associated to the outcome $x$.  
\par
The inverse of $F(\phi)$ sets a lower bound for the uncertainty
affecting the estimation of $\phi$, when a fixed observable is
measured on the system. Indeed, the FI depends on the observable that we
measure, as is apparent from the definition of $p(x \vert
\phi)$. A question naturally arises on whether there exists a
measurable observable such that the FI is maximal. Actually, this
observable always exists, though realizing it in practice can be
challenging, and the related FI is known as {\it Quantum Fisher
Information} (QFI) $H_{\phi}$ \cite{hel76,bro9x,par09}. Thus, 
we have $F(\phi) \leq H_{\phi}$
for all the possible measured observable, the QFI being defined as
\begin{equation}
H_{\phi} = \tr{\rho_{\phi}\, L_{\phi}^2}
\end{equation}
where $L_{\phi}$ is the so-called {\it Symmetric Logarithmic
Derivative Operator} (SLD operator), which is defined by
the equation $\partial_{\phi} \rho_{\phi} = \frac{1}{2}\, ( L_{\phi}
\rho_{\phi} + \rho_{\phi}L_{\phi} )$. Notice that $L_{\phi}$ is 
a self-adjoint operator with zero mean value.
Since $H_{\phi}$ maximizes the FI, from Eq.~(\ref{CRb}) we obtain the
{\it quantum} Cram\'er-Rao bound \cite{bra94,bra96}:
\begin{equation}\label{qCRb} \langle \Delta^2 \phi \rangle \geq
\frac{1}{M H_{\phi}}, \end{equation} which sets an ultimate bound for
the variance of any estimator of the parameter $\phi$, i.e. to the 
precision achievable by any inference strategy.
\subsection{QFI for Gaussian States}
Here we briefly review how to calculate the QFI and the SLD operator
for the whole class of Gaussian states \cite{zjia,amon,brau,sogeki,ban15}.
Consider a system described by a $n$-modes Gaussian
state $\rho_{\phi}$, depending on the parameter $\phi$.
Being a Gaussian state, its characteristic function can be always written as
\cite{oli:rev}:
\begin{equation*}
\chi\left[\rho_{\phi}\right](\boldsymbol{\Lambda}) = \mathrm{exp}\left\{
-\frac{1}{2}\, \boldsymbol{\Lambda}^T\, \boldsymbol{\sigma}\,
\boldsymbol{\Lambda} + i\, \boldsymbol{\Lambda}^T\, \langle
\boldsymbol{R} \rangle \right\} \end{equation*}
where $\boldsymbol{\sigma}$ is the $2n \times 2n$ real, symmetric
covariance matrix \begin{equation}\label{def_s}
\sigma_{jk} = \frac{1}{2}\,\langle R_j R_k + R_k R_j \rangle - \langle
R_j \rangle \langle R_k \rangle, \end{equation}
and $\langle \boldsymbol{R} \rangle \in \R^{2n}$ is the first-moments vector
\begin{equation}\label{def_R}
\langle R_j \rangle = \tr{\rho_{\phi}\, R_j},
\end{equation}
where we introduced the vector of canonical operators $\boldsymbol{R} =
\left( q_1,p_1,\ldots,q_n,p_n \right)$, with $[q_j, p_k] = i\, \Id\,
\delta_{jk}$.  \par
We can express the SLD operator as follows \cite{zjia}:
\begin{equation}\label{SLD}
L_{\phi} = \boldsymbol{R}^T\, \boldsymbol{\Phi}\,
\boldsymbol{R} + \boldsymbol{R}^T\, \boldsymbol{\zeta} - \nu,
\end{equation}
with a dependence at least quadratic on $\boldsymbol{R}$.
This dependence is related to the Gaussianity of the state $\rho_{\phi}$ under investigation.
Notice that $\boldsymbol{\Phi}$ is a $2n \times 2n$ real, symmetric
matrix, $\boldsymbol{\zeta}$ is a real vector of $2n$ components and $\nu$ is a scalar.
After straightforward calculation, the elements of the SLD operator $L_{\phi}$
can be linked to $\boldsymbol{\sigma}$, $\langle \boldsymbol{R} \rangle$ and
their derivatives and inverses.
In fact, we obtain that
\begin{subequations}\label{eqs}
\begin{align}
\nu &= \tr{\boldsymbol{\Omega}^T \boldsymbol{\sigma}\, \boldsymbol{\Omega}\, \boldsymbol{\Phi}}\label{eq_1}\\
\boldsymbol{\zeta} &= \boldsymbol{\Omega}^T \boldsymbol{\sigma}^{-1}\, \langle \dot{\boldsymbol{R}} \rangle\label{eq_2}\\
\dot{\boldsymbol{\sigma}} &= 2\, \boldsymbol{\sigma}\, \boldsymbol{\Omega}\, \boldsymbol{\Phi}\,
\boldsymbol{\Omega}^T \boldsymbol{\sigma} - \frac{1}{2} \boldsymbol{\Phi} \label{eq_3}
\end{align}
\end{subequations}
where $\boldsymbol{\Omega} = \overset{n}{\otimes}\, \boldsymbol{\omega}$ is the {\it symplectic matrix\/}  with $\boldsymbol{\omega} = i \boldsymbol{\sigma}_y$ and $\boldsymbol{\sigma}_y$ is the Pauli matrix.
\par
To explicitly evaluate $\boldsymbol{\Phi}$, we have to
perform a symplectic diagonalization of the 
covariance matrix $\boldsymbol{\sigma}$.
Thus, we define $\boldsymbol{\sigma}_S = S\, \boldsymbol{\sigma}\, S^T$ the
diagonalized 
covariance matrix, where $S$ is a suitable symplectic transformation,
$S\, \boldsymbol{\Omega}\, S^T = \boldsymbol{\Omega}$.
Therefore, we obtain
\begin{equation}
\left(\boldsymbol{\Phi}_S\right)_{jk} = \frac{\left(\boldsymbol{\Omega}^T\, \boldsymbol{\sigma}_S\,
\dot{\boldsymbol{\sigma}}_S\, \boldsymbol{\sigma}_S\, \boldsymbol{\Omega} + \frac{1}{4}\, \dot{\boldsymbol{\sigma}}_S
\right)_{jk}}{2\, \lambda_j^2\, \lambda_k^2 - \frac{1}{8}}
\end{equation}
where $\boldsymbol{\Phi}_S = S\, \boldsymbol{\Phi}\, S^T$ and $\lambda_j$
are the eigenvalues of $\boldsymbol{\sigma}$.
Eventually, we obtain the matrix $\boldsymbol{\Phi}$ by applying
the inverse of the symplectic transformation $S$ and we can evaluate the SLD
operator and, thus, the QFI for a generic Gaussian state.
In particular, in the case of Gaussian pure states
we have $\lambda_j = 1/2$, $\forall j$ and an explicit
equation for $\boldsymbol{\Phi}$ can be written:
\begin{subequations}\label{phi_pur}
\begin{align}
\boldsymbol{\Phi} &= \frac{1}{4}\, \boldsymbol{\Omega}^T\, \boldsymbol{\sigma}^{-1}\,
\dot{\boldsymbol{\sigma}}\, \boldsymbol{\sigma}^{-1}\, \boldsymbol{\Omega} \\
&= - \dot{\boldsymbol{\sigma}}
\end{align}
\end{subequations}
\par
Summarizing, from Eq.~(\ref{SLD}) we can obtain the the following
expression for the QFI: \begin{equation}\label{QFIG}
H_{\theta} = \tr{\boldsymbol{\Omega}^T\, \dot{\boldsymbol{\sigma}}\, \boldsymbol{\Omega}\, \boldsymbol{\Phi}}
+ \langle \dot{\boldsymbol{R}} \rangle^T \boldsymbol{\sigma}^{-1} \langle \dot{\boldsymbol{R}} \rangle
\end{equation}
which holds for pure and mixed Gaussian states.
\subsection{Sensitivity and FI}
Here we introduce another quantity, known as {\it sensitivity},
related to the precision of the estimation of an unknown parameter $\phi$
once a measurement is chosen.
\par
We assume that an observable $X$ is measured on the system
under examination, described by $\rho_{\phi}$, and that the mean value
$\langle X \rangle \equiv X(\phi)$ depends on the parameter $\phi$.
If $\phi$ is shifted by a quantity $\delta \phi \ll 1$,
the mean value, up to first order in $\delta \phi$ is now given by
\begin{equation}\label{var_d}
X (\phi + \delta \phi) \approx X (\phi)
+ \delta \phi\, \partial_\phi X(\phi)\,,
\end{equation}
i.e. $X$ is shifted by a quantity (we drop the explicit
dependence on $\phi$):
\begin{equation}
\Delta  X  = \delta \phi\, \partial_\phi X.
\end{equation}
Now, if we want to detect any shift of the order $\delta \phi$ by
looking at $X$, the absolute value of its variation $|\Delta X| = 
|X (\phi + \delta \phi) - X (\phi)|$ has to be larger than the 
statistical fluctuations of the mean value itself, i.e. the 
square root of the variance
$\sqrt{\Delta X^2} = \sqrt{\langle X^2\rangle - \langle X \rangle}$.
Otherwise, we would not be able to
say whether the change of $X$ was due to random fluctuations,
or to an actual shift of $\phi$.
\par
At least, the uncertainty $\sqrt{\Delta X^2}$ and its variation
$| \Delta X |$ can be equal. From this request we
obtain the minimum value of $\delta \phi$ that can be sensed
by looking at changes in the values of $X$. Such minimum value 
is called sensitivity, and it is expressed as:
\begin{equation}\label{sens}
S_{\phi} \equiv \frac{\sqrt{
\Delta X^2 }}{\left| \partial_\phi X \right|}.
\end{equation}
It is clear that $S_\phi$ may depend on $\phi$ and, in general, 
it is minimized for a particular choice $\phi=\phi_0$ that is 
usually referred to as an optimal {\it working point}.
\par
It is worth noting that, in this setup, both the state of the
system $\rho_{\phi}$ and the observable $X$ are given, 
and therefore the FI may, at least in principle, be 
obtained. However, it is often impossible to obtain the 
analytic form of the FI since evaluating the probability 
distribution $p(x \vert \phi)$ may be 
challenging. Moreover, it should be noticed that under specific 
conditions, the FI and the sensitivity are equal, thus further
motivating the use of the latter in place of the FI.
Indeed, whenever a Gaussian approximation of the distribution 
$p(x \vert \phi)$ may be assumed, with the mean value $X$
and variance $\sigma^2 = \Delta X^2$ depending
on the parameter $\phi$, then, from Eq.~(\ref{FI}) follows that:
\begin{equation}
F(\phi) = \frac{\left( \partial_\phi \, X \right)^2
+ 2 \left( \partial_\phi	\sigma \right)^2}{\sigma^2}.
\end{equation}
If now $\sigma$ slowly changes at the working point,
namely, $\partial_\phi\sigma|_{\phi = \phi_0}\approx 0$, 
we obtain $S_{\phi}^2 = 1/F(\phi)$, while in general 
$S_{\phi}^2 \geq 1/F(\phi)$, as it should be since the sensitivity
is built to assesses the precision of an estimator based on the sole
mean value of the distribution.
\section{Passive and active Interferometers}\label{s3:api}
\begin{figure}[h!]
\center
\includegraphics[width=0.9\columnwidth]{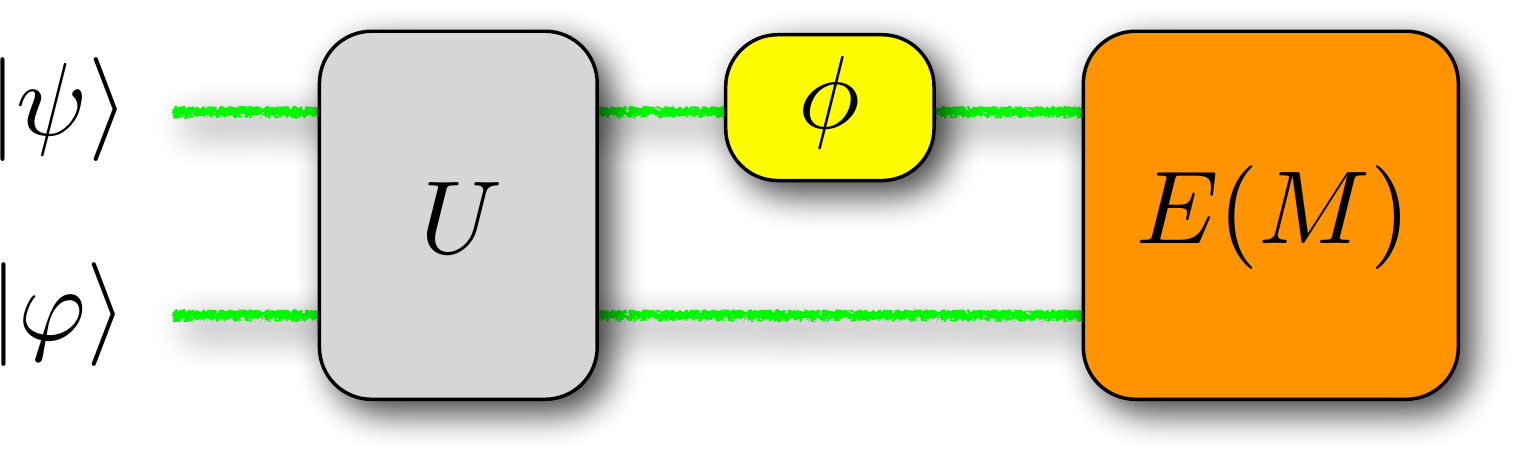}
\includegraphics[width=0.9\columnwidth]{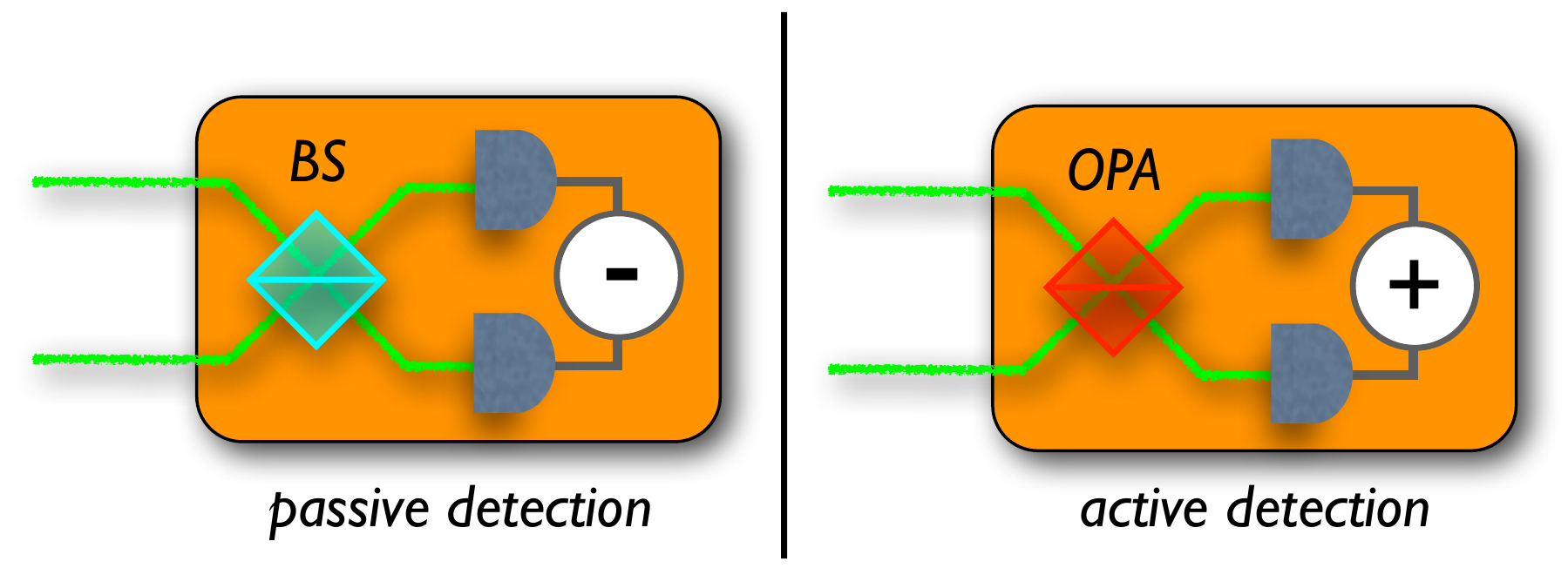}
\caption{(Color online) The upper panel shows the general scheme 
of an interferometric setup. A factorized pure state 
$\ket{\psi}\ket{\varphi}$ of the two modes is injected in the device.
The two modes are then coupled by the unitary operator $U$ 
and a unknown phase shift $V(\phi)$
is applied to one of the beams. At the of output the interferometer, 
an observable described by the POVM $E$ is measured. The lower left
panel shows a schematic diagram of the passive measurement stage.
Before being measured, the two beams go through a beam splitter. 
The difference photocurrent $D_{-}$ is then measured. 
In the lower right panel: scheme of the active measurement
stage. The beam splitter is here replaced by an OPA. The 
measured observable is the sum photocurrent $D_{+}$.}
\label{f1:sch}
\end{figure}

A general interferometric scheme may be sketched as in 
the upper panel of 
Fig.~\ref{f1:sch}. Two radiation beams are injected into 
the interferometer in a factorized state 
$|\Psi_\ip\rangle\rangle= \ket{\psi}\otimes \ket{\varphi}$
that we will assume to be pure and Gaussian. 
The first stage consists of a unitary  
operation $U$ that couples the two beams, followed by a phase shift 
$V(\phi)= e^{-i a^\dag a \phi} \otimes {\Bbb I}$ that is applied 
to one of the two arms. At the output, an observable described 
by the POVM $E$ is measured on the whole system, with the aim, 
of inferring the value of $\phi$ after a suitable data processing.
\par
In this paper we are going to consider two different classes of
interferometers. In the first class we have devices employing {\it 
active components}, such as optical parametric amplifiers (OPAs) 
\cite{agar:10}. The second one, instead, includes devices
with {\it passive components}, e.g. beam splitters.
Active components differ from passive ones since they increase 
the energy of the incoming light beams, while the passive ones 
keep it constant. The action of an OPA on a two-mode state is 
described by the unitary operator 
$$U_{\OPA}(\zeta) = \mathrm{exp}\{\zeta\, a^{\dagger}b^{\dagger}
-\zeta^{\ast}\, a\, b\},$$ 
where $a$ and $b$ are the two field operators describing
modes, and $\zeta \in \C$ is a coupling constant, linked to 
the gain of the amplifier. A beam splitter is instead described 
by the unitary 
$$U_{\BS}(\nu) =
\mathrm{exp}\{\nu\, a^{\dagger} b -\nu^{\ast}\, a\, b^{\dagger}\},$$
where  $\nu \in \C$ is a coupling constant which 
determines the transmissivity of the beam splitter.
\par
The performances of both classes of interferometers may be assessed
using quantum estimation theory , which provides tools to find the
optimal working regimes, i.e. the optimal input signals and the optimal
detection stage (see Sect.~\ref{s4:qfii})\cite{CS:JOSAB}.  However, the
realization of the optimal detection stage is usually challenging with
current technology and thus it becomes relevant to assess the precision
achievable by feasible schemes. In this paper we consider two specific
measurement schemes, characterized by their
passive/active nature, optimizing their performances over the input
signals in different configurations. 
\par
In the passive detection scheme, the two radiation beams interferes at a 
beam splitter and then are detected by two photodetectors, 
which count the number of photons. In this case, the measured 
observable is the {\it difference photocurrent} $$D_{-} = a^{\dagger} a - 
b^{\dagger} b$$ between the signals, i.e. the state just before the
detectors
\begin{align}
|\Psi_\ot\rangle\rangle = U_{\BS}^\dag (\nu) V(\phi) U
\, |\Psi_\ip\rangle\rangle\,.
\end{align}
In the active configuration, the beams splitter is 
replaced by an OPA and the measured 
observable is now the {\it sum photocurrent}
$$D_{+} = a^{\dagger} a + b^{\dagger} b,$$ i.e. the total number of 
photon  of the output signals, which, in this case, are described 
by the state 
\begin{align}
|\Psi_\ot\rangle\rangle = U_{\OPA}^\dag (\zeta) V(\phi) U
|\Psi_\ip\rangle\rangle
\end{align}
The two possible detection stages are schematically depicted in the
lower panel of Fig. \ref{f1:sch}.
\par
\section{QFI for passive and active interferometers}\label{s4:qfii}
Here we briefly review the optimal performances achievable
by passive and active interferometers \cite{CS:JOSAB} with Gaussian
input signals. Results will serve as a referece to assess the 
performances of the four concrete configurations analyzed in the 
following Section.
\subsection{Passive quantum interferometer}\label{s4.1:qfipi}
The scheme of a typical passive interferometer is described in
Fig.~\ref{f1:sch}, where the unitary operator $U$ represents a 50:50 beam
splitter, and the input states $\ket{\psi}\ket{\phi}$ are assumed to be
two displaced-squeezed states. The QFI is, by definition, optimized over
all the possible measurements, and therefore we do not define any
measurement stage.  The two input states can be written as $\ket{\alpha,
\xi}\ket{\gamma, \zeta}$, where $\alpha, \gamma \in \R$ are the coherent
coefficients (no phase is considered), and $\xi \in \R^+$, $\zeta \in
\C$ are the squeezing coefficients, where the complex phase is accounted
only for the second one.  The parameter $\zeta$ can be decomposed as
$\zeta = r e^{-i \theta}$, where $r \in \R^+$ is its modulus, while
$\theta \in [0, 2 \pi)$ is the phase.  The QFI $H_\phi$ may be 
evaluated using Eq.~(\ref{QFIG}), starting from the first-moment vector
and the covariance matrix after the beam splitter, we have  
\begin{align}\nonumber
H_{\phi}= &\frac{1}{4}\, \Big\{4\, e^{2 \xi}\, 
(\alpha +\gamma )^2 + \cosh 4 \xi +2 \cos \theta \sinh 2 r \\ \nonumber
&\times \left( 2\, (\alpha +\gamma)^2 + \sinh 2 \xi \right) + 4\, 
(\alpha +\gamma)^2 \cosh 2 r \\
&+ \cosh 2 (r - \xi) +\cosh 2 (\xi + r) + \cosh 4 r - 4 \Big\}
\label{QFI_MZ}
\end{align}
which depends on the coherent amplitudes and the squeezing parameters, 
while it is independent on the phase-shift itself, being the problem
covariant.
\par
As a matter of fact, there are several parameters involved 
in  the optimization and thus it is useful to introduce a
suitable re-parameterization to emphasize the quantities that are
physicaly relevant. In particular, we are interested in the
behavior of the QFI $H_{\phi}$ as a function of the overall 
intensity (i.e., the total number of photons) of the input 
beams. The first parameter we introduce is the 
{\it coherent trade-off coefficient}
$\delta = \alpha^2/(\alpha^2 + \gamma^2)$, $\delta \in [0,1]$ ,
assessing the fraction of coherent photons in each input beams.
Then, we denote by $N_{\mathrm{tot}}$ the {\it total 
average number of photons} of the system, accounting for 
both coherent and squeezed
photons, as $N_{\mathrm{tot}} =  \alpha^2 + \gamma^2 + \sinh^2 \xi +
\sinh^2 r$.  To describe the squeezing properties of the beams, we use
the parameters $\beta_{\mathrm{tot}}$ and $\beta$. The first one is the
fraction of (total) squeezed photons, namely
$\beta_{\mathrm{tot}} = (\sinh^2 \xi
+ \sinh^2 r)/N_{\mathrm{tot}}$, $\beta_{\mathrm{tot}} \in [0,1]$.
The second one is the ratio between the average number of
squeezed photons in one branch and the total average number of photons,
namely, $\beta = \sinh^2 \xi / N_{\mathrm{tot}}$, $\beta \in [0,
\beta_{\mathrm{tot}}]$.  
\begin{figure}[h!]
\includegraphics[width=0.48\columnwidth]{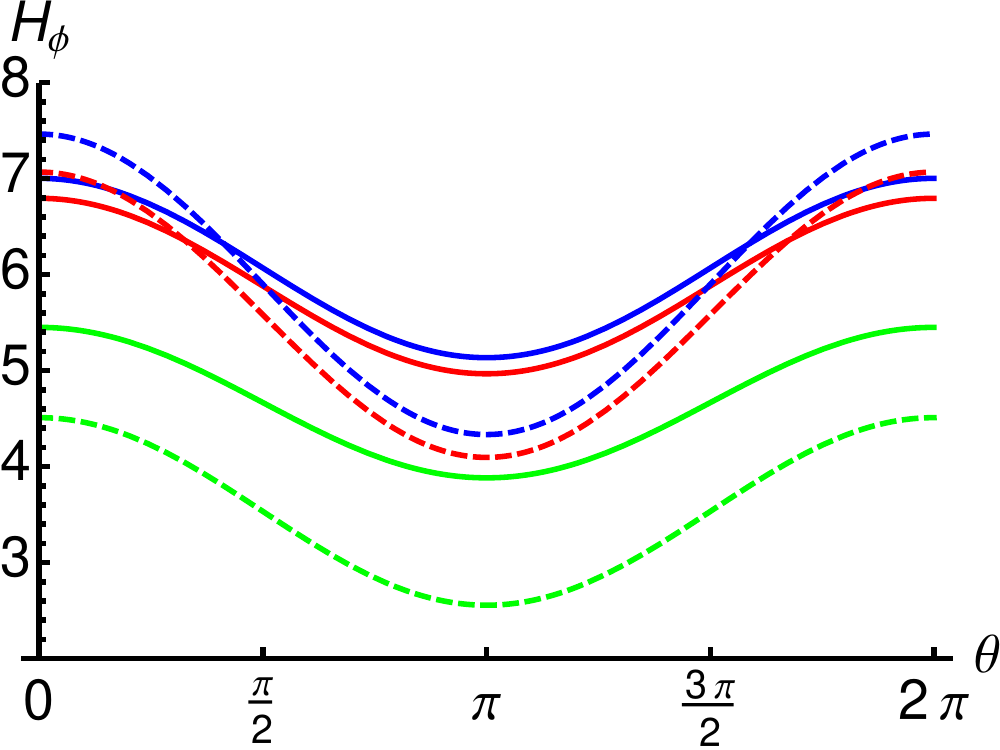}
\includegraphics[width=0.48\columnwidth]{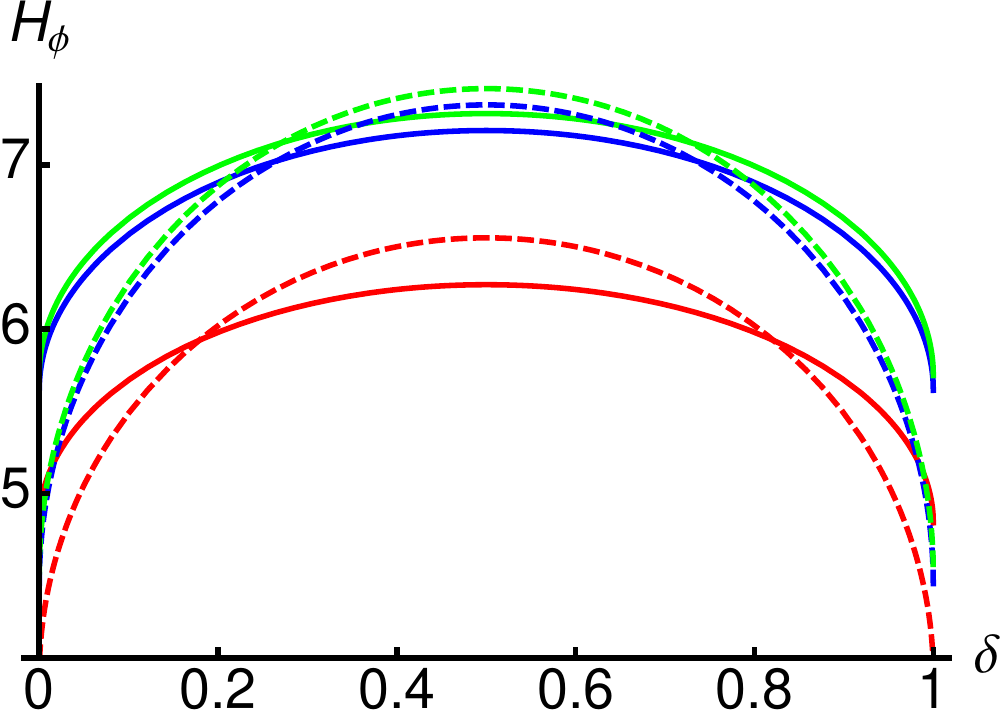}
\includegraphics[width=0.48\columnwidth]{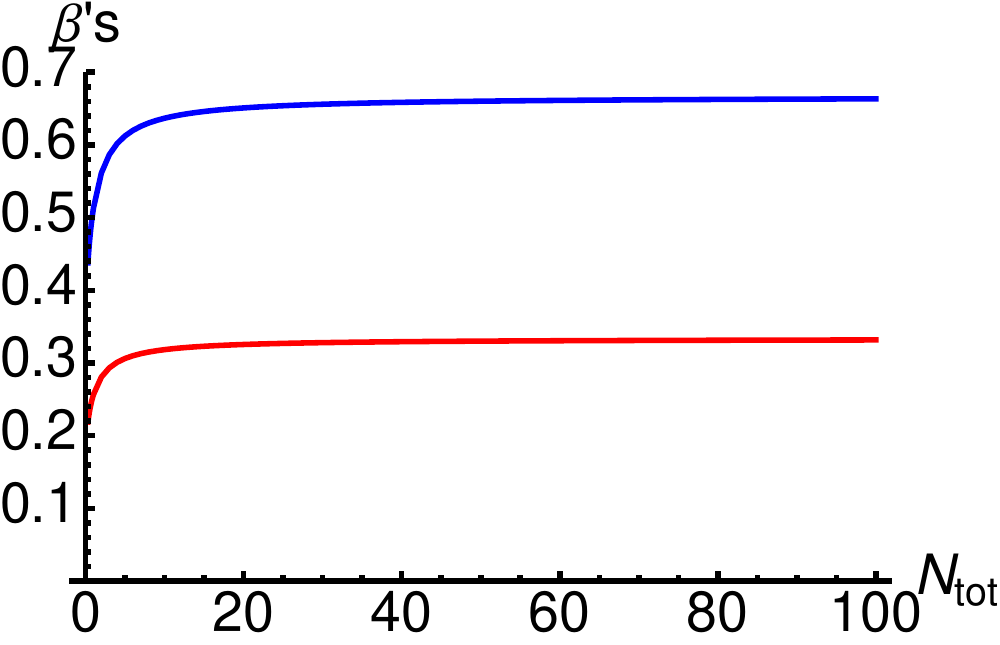}
\includegraphics[width=0.48\columnwidth]{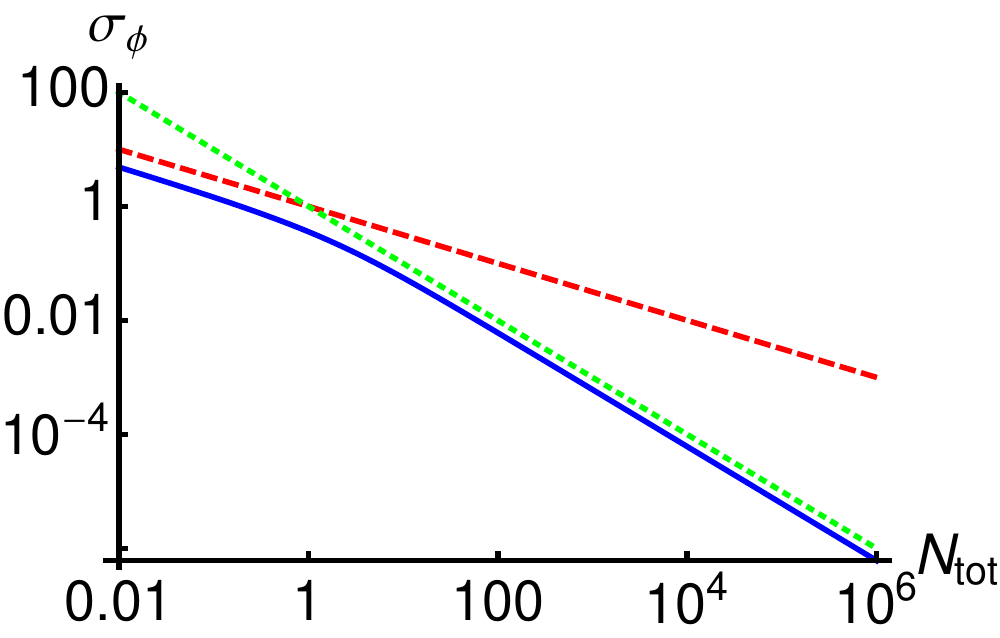}
\caption{(Color online) (upper left) The $H_{\phi}$ as a
function of $\theta$, where $N_{\mathrm{tot}} = 1$, and $\delta = 1/4$
(red lines), $\delta = 1/2$ (blue lines), and $\delta = 1$ (green
lines).  Solid lines for $\beta_{\mathrm{tot}} = 3/4$ and $\beta = 2/3$,
and dashed lines for $\beta_{\mathrm{tot}} = 1/3$ and $\beta = 1/6$.
(upper right) The $H_{\phi}$ as a function of $\delta$, where
$N_{\mathrm{tot}} = 1$, and $\theta = 0$. Solid lines for
$\beta_{\mathrm{tot}} = 3/4$, dashed lines for $\beta_{\mathrm{tot}} =
1/3$, and red lines for $\beta = 0$, blue lines for  $\beta =
\beta_{\mathrm{tot}}/4$, and green lines for $\beta =
\beta_{\mathrm{tot}}/2$.  (lower left) The values of $\beta_{\mathrm{tot}}$
(blue line), and $\beta$ (red line) maximizing the Quantum Fisher
Information $H_{\phi}$, as a function of $N_{\mathrm{tot}}$.
(lower right)
The standard deviation $\sigma_{\phi}$ (blue solid line),
as a function of $N_{\mathrm{tot}}$. 
The shot-noise limit $1/\sqrt{N_{\mathrm{tot}}}$ is the red dashed line,
and the Heisenberg limit $1/N_{\mathrm{tot}}$ is the green dotted line.} 
\label{f2:pas}
\end{figure}
\par 
In Fig.~\ref{f2:pas} 
we illustrate the features of the QFI by
showing its behavior as a function of a given parameter and by 
fixing the others. Looking at the upper left panel 
we see that $H_{\phi}$ is maximized by choosing the squeezing phase 
$\theta$ of equal to $0$ (or $2\pi$) independently on the value of the
other parameters (while the actual value of the
maximum depends on the other involved parameters). This means that
optimal input signals should have the same squeezing phase, i.e. 
squeezing may be chosen real without loss of generality (or more
generally, in phase with the coherent amplitudes).
Concerning the dependence on $\delta$, the upper right panel of 
Fig.~\ref{f2:pas} shows that
$H_{\phi}$ is maximized by $\delta = 1/2$ independently on the other 
parameters, i.e  in the optimal input signals the
average number of the coherent photons in the two input states should be
the same. Finally, in the lower left panel of  Fig.~\ref{f2:pas}, we 
show the optimal values of $\beta_{\mathrm{tot}}$ (blue line) 
and $\beta$ (red line) maximizing $H_{\phi}$. Interestingly, the 
optimal $\beta$ is equal to
$\beta_{\mathrm{tot}}/2$ and, in the limit $N_{\mathrm{tot}} \gg 1$,
$\beta_{\mathrm{tot}} \to 2/3$. Therefore, we conclude that in the
optimal case the squeezing has to be balanced between the two input
states, and we also need a given number of coherent photons.
\par
Overall, the optimization reveals that the best input signals 
correspond to two identical displaced-squeezed states
$\ket{\alpha, r}$.  It is worth noting that the state of the system
after the beam splitter is factorised, and is equal to $\ket{\sqrt{2}
\alpha, r} \ket{0, r}$, where $\ket{\sqrt{2}\alpha, r}$ undergoes a
phase shift, while $\ket{0, r}$ plays the role of a quantum-enhanced 
phase reference \cite{tra11}. The maximized QFI may be written as 
\cite{CS:JOSAB}:
\begin{equation*}
H_{\phi}^{\rm max}(N_{\mathrm{tot}}) = \frac{4 N_{\mathrm{tot}}}{9} 
\left[ 2\, \sqrt{N_{\mathrm{tot}} (N_{\mathrm{tot}} + 3) } + 4\, N_{\mathrm{tot}} + 9\right],
\end{equation*}
which in the high-energy limit ($N_{\mathrm{tot}} \gg 1$) reduces to
$H_{\phi}^{\rm max}(N_{\mathrm{tot}}) \approx 8/3\, (N_{\mathrm{tot}}^2 + 2\, N_{\mathrm{tot}})$.
\par
The minimum detectable fluctuation of the phase $\phi$ can now be
obtained using the Cram\'er-Rao bound, Eq.~(\ref{CRb}).  In
the lower right panel of Fig.~\ref{f2:pas} we show the behaviour 
of $\sigma_\phi =
1/H_\phi$, i.e. the minimum detectable fluctuation of $\phi$, as 
a function of the total energy. We have employed quantum states of light, and 
indeed the sensitivity is enhanced compared to the shot-noise limit, 
achieving the Heisenberg scaling. Moreover, having maximized the QFI 
in the most general case (for a passive device), we have found the 
actual ultimate limit for the precision of this kind of interferometer.
\subsection{Active quantum interferometer}\label{s4.2:qfiai}
We consider active interferometer that employs an OPA in 
place of the beam splitter. The scheme is 
shown in Fig.~\ref{f1:sch}, where the unitary operator represents 
the action of the OPA. Since the quantumness needed to 
beat the shot-noise limit is now provided by the OPA, we are 
led to consider just coherent states as input signals. Besides,
we assume that the two coherent amplitudes are real
namely, $\ket{\alpha}\ket{\gamma}$, with $\alpha, \gamma \in \R$, 
i.e. the two signals have the same phase. As we will see in the 
following, this choice lead to Heisenberg scaling of
sensitivity.
The unitary operator describing the amplifier is $U_{\rm OPA}(\zeta)$,
where $\zeta = r\, e^{-i \theta}$ is the squeezing coefficient,
$r \in \R^+$ and $\theta \in [0, 2 \pi)$.
Upon using again Eq.~(\ref{QFIG}), starting from the first-moment vector
and the covariance matrix after the OPA, we have  
\begin{align}\nonumber
H_{\phi}= & \alpha^2 + \gamma^2 + \left( \alpha^2 + 
\gamma^2 + \frac{1}{2} \right) \cosh 4 r\\
&+ 2\, \alpha \gamma \cos \theta \sinh 4 r + 2\, 
(\alpha^2 - \gamma^2) \cosh 2 r - \frac{1}{2}\label{QFI_act}
\end{align}
As in the case of passive interferometers, we want to write the QFI as a
function of suitable parameters, related to the energetic properties of
the light beams.  To this aim, we still use the coherent trade-off
coefficient $\delta \in [0, 1]$, introduced above, and the total average
number of photons (including those introduced with the OPA)
$N_{\mathrm{tot}} = (\alpha^2 + \gamma^2 + 1) \cosh 2 r + 4\, \alpha
\gamma \cos \theta \sinh r \cosh r - 1$, $N_{\mathrm{tot}} \in [0,
\infty]$.  The last parameter we define is the ratio between the number
of squeezed photons and the total number of photons, namely, $\beta =
2\, \sinh^2 r / N_{\mathrm{tot}}$, $\beta \in [0,1]$.
\par
\begin{figure}[h!]
\includegraphics[width=0.48\columnwidth]{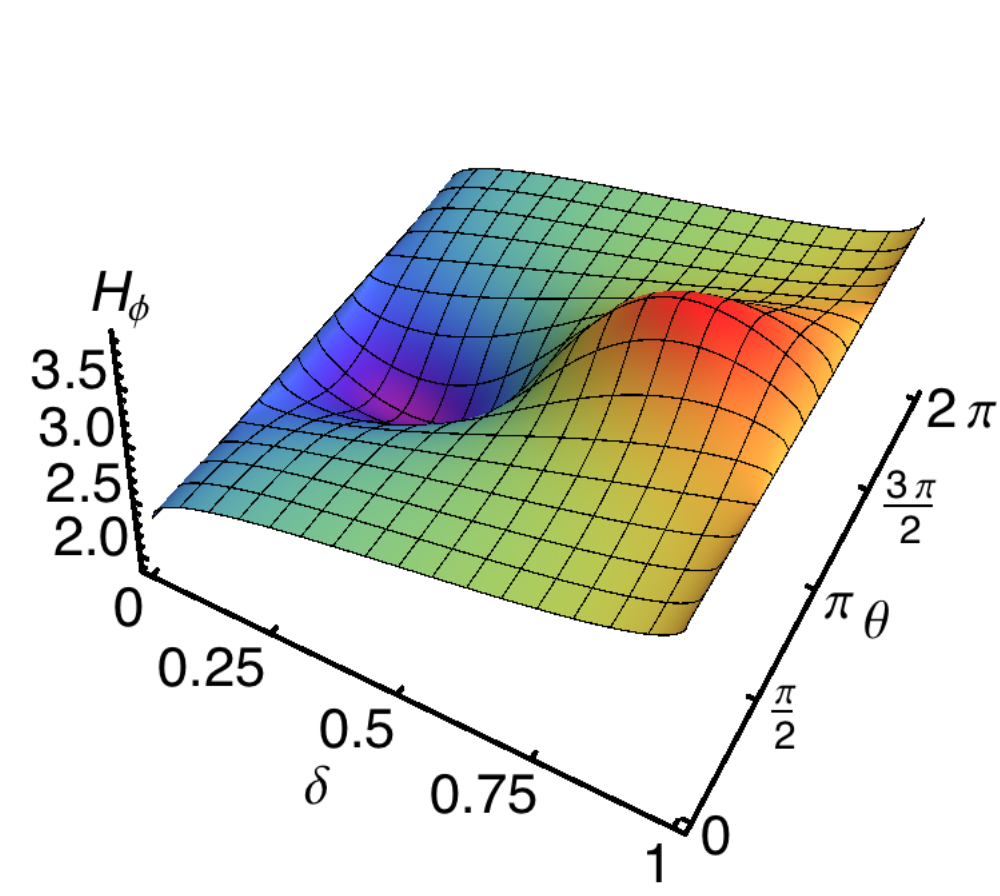}
\includegraphics[width=0.50\columnwidth]{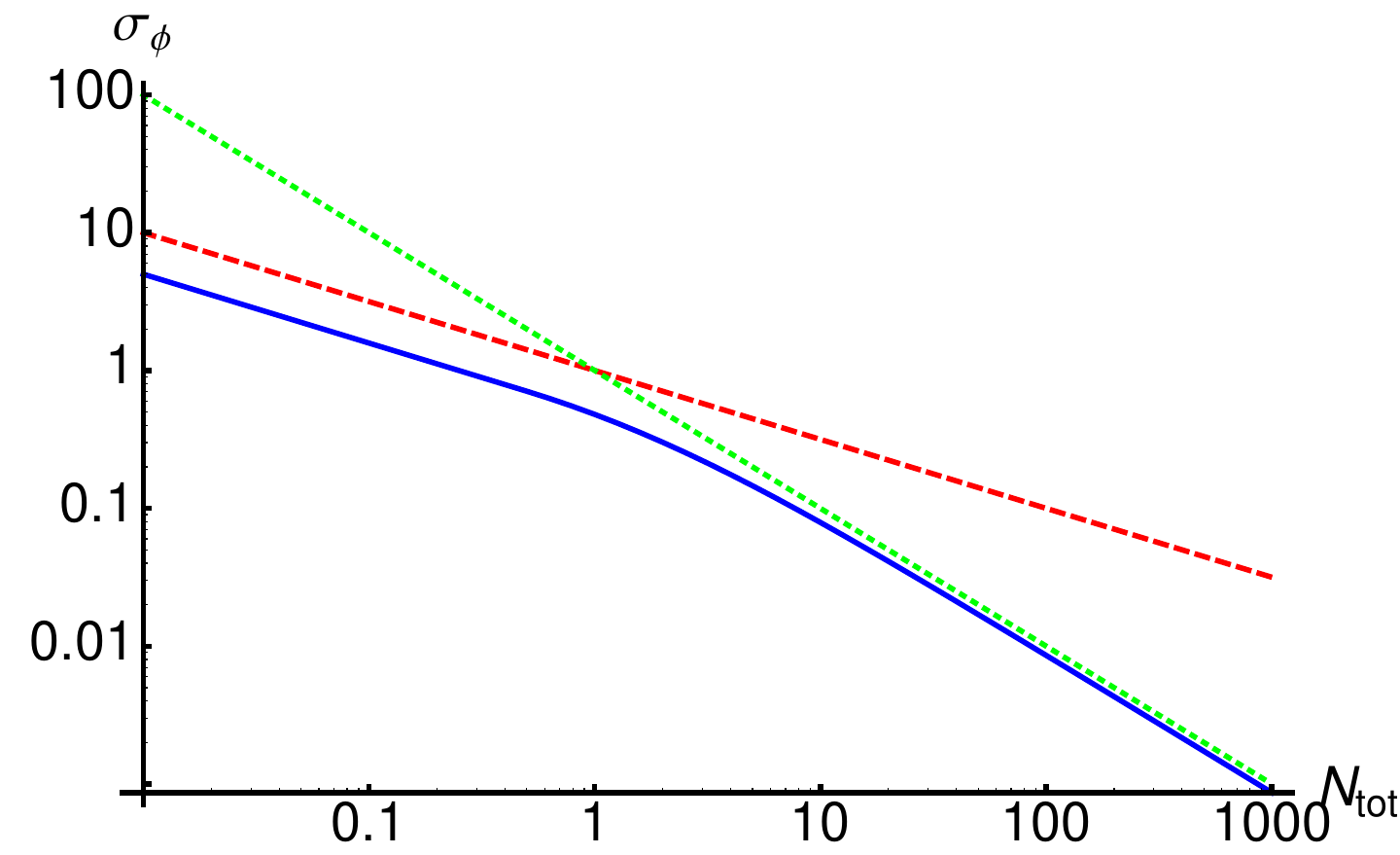}
\center
\includegraphics[width=0.8\columnwidth]{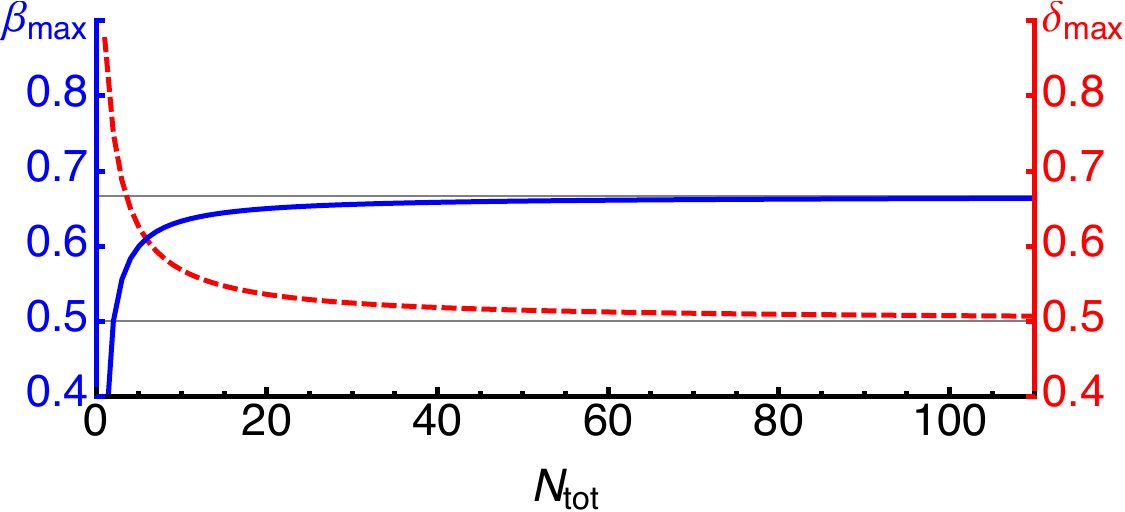}
\caption{(Color Online)
(Upper left)
The QFI  $H_{\phi}$ for an active interferometer as a function
of $\theta$ and $\delta$, for fixed values of $\beta = 0.7$
and $N_{\mathrm{tot}} = 1$.
(Upper right) The minimum detectable fluctuation $\sigma_{\phi}$ (blue solid
line) as a function of $N_{\mathrm{tot}}$, with the shot-noise limit
(red dashed line) and the Heisenberg limit (green dotted line).
(Lower panel) The parameters $\beta_{\mathrm{max}}$ and
$\delta_{\mathrm{max}}$, maximising $H_{\phi}$, as a function of
$N_{\mathrm{tot}}$.}
\label{f3:act}
\end{figure}

In order to maximize the QFI $H_\phi$, we analyze its behavior as 
a function of $\theta$ and $\delta$ for fixed $\beta$ and 
$N_{\mathrm{tot}}$. The typical behaviour is shown in the upper left
panel of Fig. \ref{f3:act}. As a matter of fact, the  
maximum is achieved for $\theta = \pi$ independently on the other
parameters, while the optimal value of
$\delta$ depends on $\beta$ and $N_{\mathrm{tot}}$ (we found that 
$\delta > 1/2$). Given 
$N_{\mathrm{tot}}$, in order to find the values of
$\beta$ and $\delta$ maximizing the QFI we should use a numerical
maximization.  The results are shown in the upper right panel of
Fig.~\ref{f3:act}, where the minimum detectable fluctuation of the
phase $\phi$ is plotted.
Remarkably, also with an active interferometer the Heisenberg limit can
be achieved \cite{agar:10,CS:JOSAB}.  In  the lower panel of
Fig.~\ref{f3:act}, we show the values of $\delta_{\mathrm{max}}$
and $\beta_{\mathrm{max}}$ maximising  $H_{\phi}$ as functions of
$N_{\mathrm{tot}}$. In the high-energy case ($N_{\mathrm{tot}} \gg 1$)
the parameter $\delta_{\mathrm{max}}$ is equal to $1/2$, while
$\beta_{\mathrm{max}}$ is equal to $2/3$ leading to the following
analytic expression for the QFI: 
\begin{equation*}
H^{\rm max}(N_{\mathrm{tot}}) \approx
\frac{4}{3}\, \left(N_{\mathrm{tot}}^2 + 2 N_{\mathrm{tot}}\right)
\qquad (N_{\rm tot} \gg 1).
\end{equation*}
It is worth noting that in order to achieve this value, two coherent
states with the same number of photons have to be injected in the
interferometer ($\delta_{\mathrm{max}} \simeq \frac12$ for large
$N_{\mathrm{tot}} \gg 1$), and the OPA has to introduce two third 
of the total
number of photons in the system. Therefore, in the regime $
N_{\mathrm{tot}} \gg 1$ the ultimate limit of the QFI for active
interferometers is proportional to the square of the total number of
photons in the system.
\par
Summarizing, the precision achievable by
active interferometers shows the same scaling of 
passive ones. However, passive devices offers a factor 
two enhacement over active ones and should thus be 
preferred, assuming that their
implementations involve similar technological efforts. We notice 
that both schemes allows one to beat the classical
precision limit, upon the introducti of  squeezed light in the system.
\section{Sensitivity for passive and active interferometers}\label{s5:sapi}
In the previous Section we evaluated the ultimate limits to 
precision for any phase-shift estimation scheme based on passive
and active interferometers. The Cramer-Rao theorem ensures that
the obtained bounds are achievable, i.e. there exists an observable
which may be employed to estimate the phase-shift with optimal
precision. This optimal observable, however, correspond to the spectral
measure of the SLD \cite{hel76,par09} and it is not clear whether, and
in which regimes, it may be implemented with current optical technology.  
\par
In order to assess the performances of feasible interferometers
in this section we consider a realistic detection stage
and analyze the sensitivity of different configurations, also taking
into account possible imperfections of the detectors, such as losses
leading to non unit quantum efficiency.
In particular, we evaluate phase sensitivity using Eq.~(\ref{sens}), 
with the observable $X$ replaced by either the difference 
photocurrent $D_{-}(\eta)$
in the passive measurement scheme, or by the sum photocurrent
$D_{+}(\eta)$ in the active case, see Appendix \ref{a1:mea} for 
the expression of the mean values and the variances. 
We optimize the input signals 
in all the four possible configurations (active/passive interferometers
with active/passive detection stage) and compare performances in 
the ideal case, as well as for non unit quantum efficiency.
\subsection{The passive/passive case}
\label{ss:PassIntPassMeas}
We now consider the passive interferometer introduced in
Sec.~\ref{s4.1:qfipi}, equipped with the passive measurement scheme
described in the lower left panel of Fig.~\ref{f1:sch}.  This kind of
device is the well-known Mach-Zehnder interferometer, also 
equivalent to the Michelson one employed in
gravitational interferometers \cite{pez08,grav:11,raf14,lan13,lan14}.  It
possible to show \cite{oli:par:OptSp} that in this case the best choice
for the input states is $\ket{\alpha, \xi}\ket{0, \zeta}$, where
$\alpha, \xi \in \R$ are, respectively, the coherent and squeezed
coefficients of the first state and $\zeta = r\, e^{-i \theta}$ is the
squeezing coefficient of the second state, $r \in \R^+$ and $\theta \in
[0,2\, \pi)$.  Furthermore, both the beam splitters used in the
interferometer are balanced, and their phases differ by $\pi$, so that
the output is equal to the input if
there is no additional phase shift.
\subsubsection{Ideal photodetection}
\begin{figure}[h!]
\includegraphics[width=0.49\columnwidth]{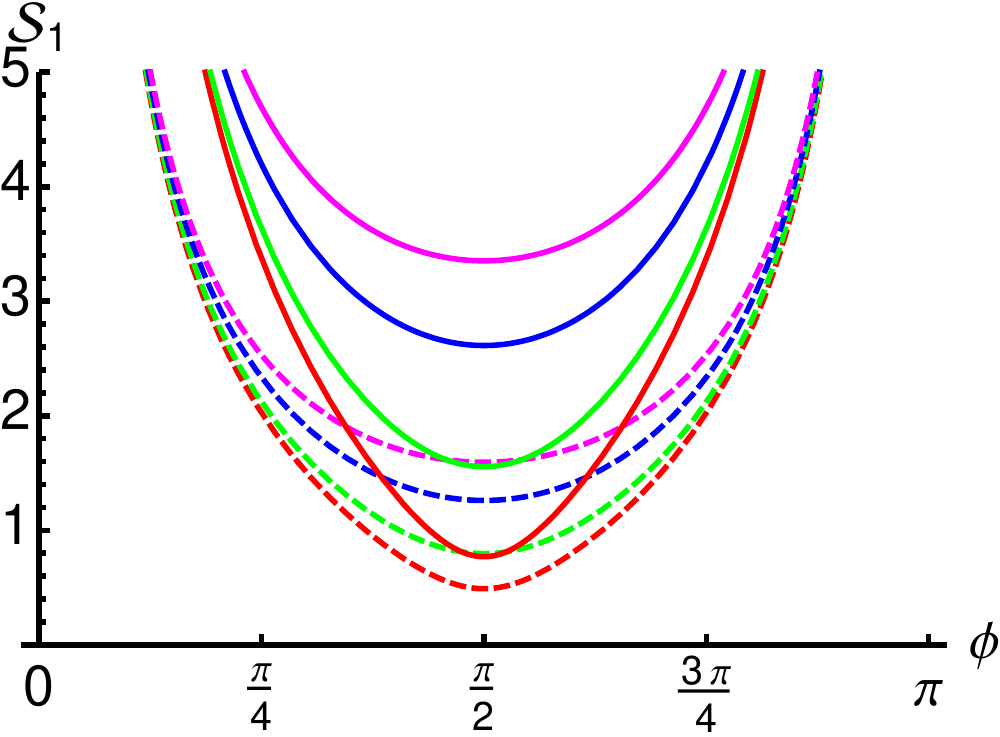}
\includegraphics[width=0.49\columnwidth]{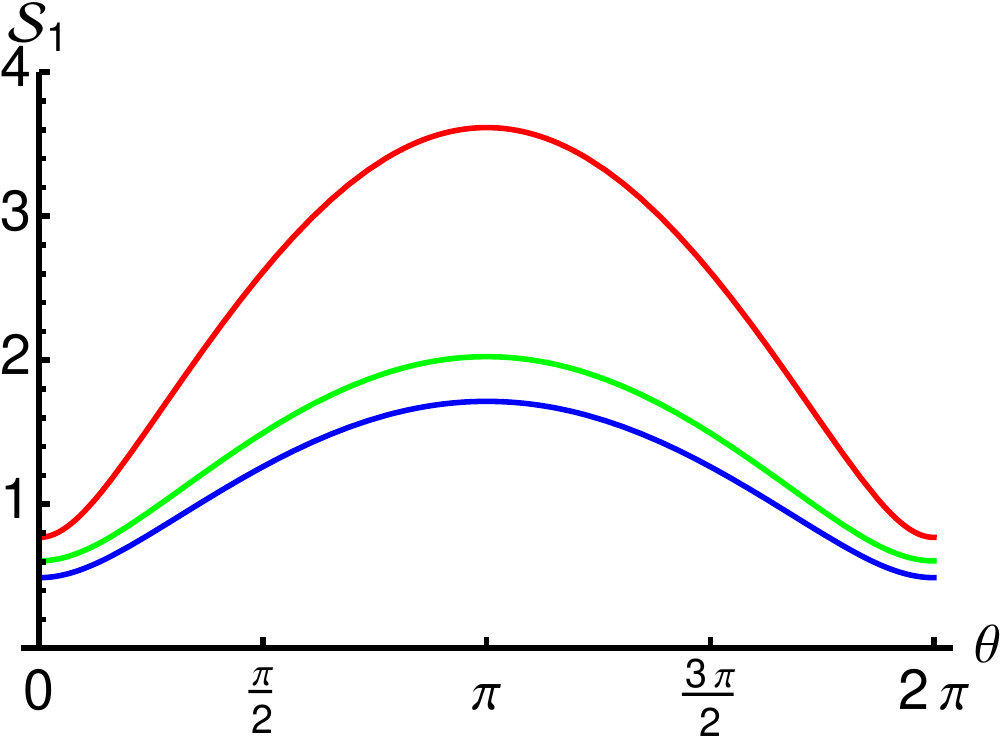}
\includegraphics[width=0.49\columnwidth]{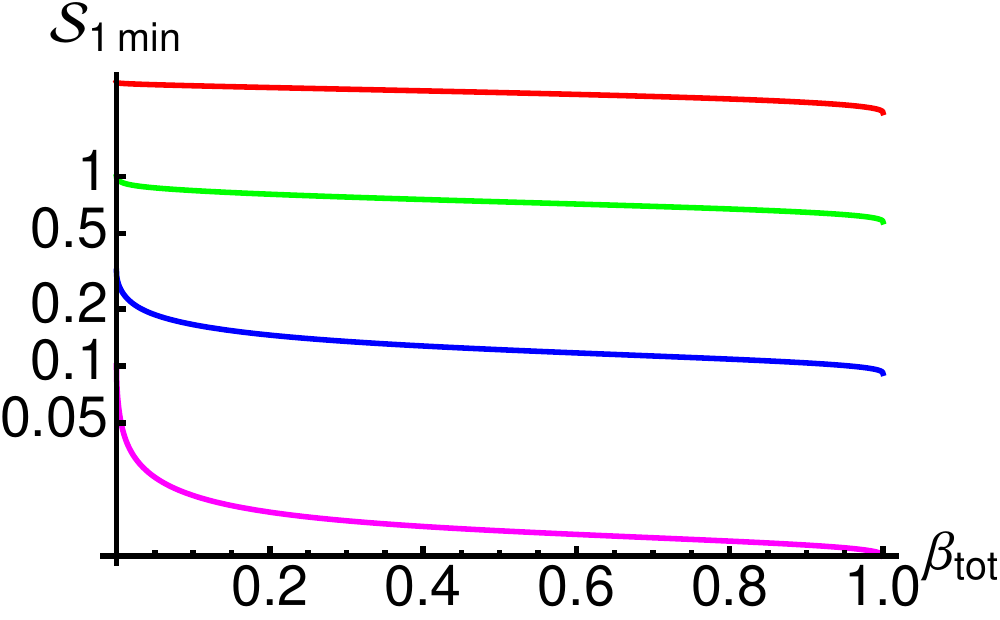}
\includegraphics[width=0.49\columnwidth]{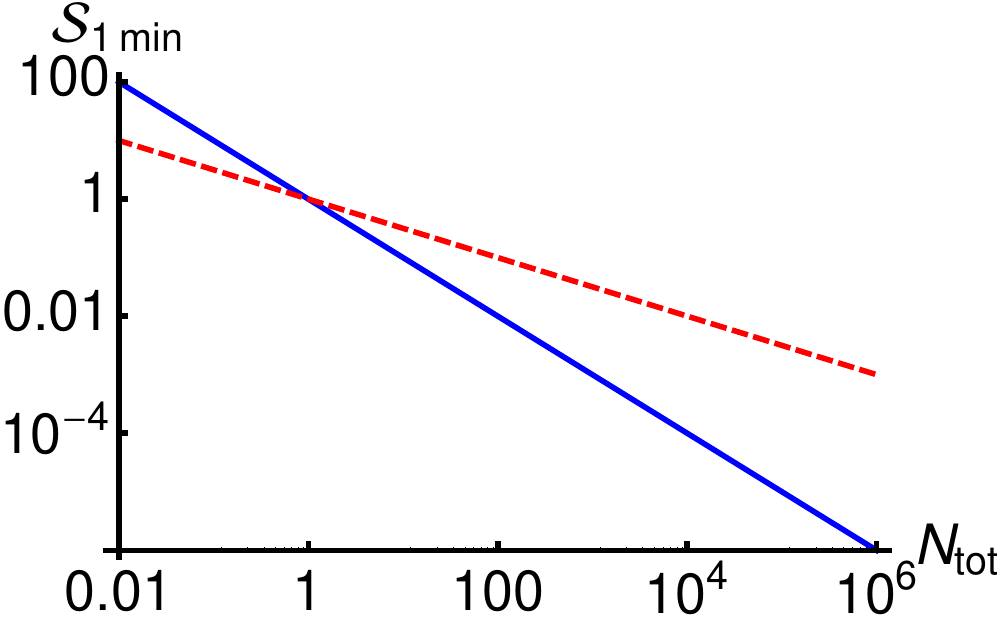}
\caption{(Color Online)
The passive/passive interferometer with ideal phodetection.
In the upper left panel we show the sensitivity 
$S_1$ as a function of $\phi$, for
$N_{\mathrm{tot}} = 2$, $\xi = 0.5$ and $r = 0.7$ (solid line), or $\xi
= 0.7$ and $r = 0.5$ (dashed line), and for $\theta = 0$ (red lines), or
$\pi/4$ (green), or $\pi/2$ (blue), or $3/4\, \pi$ (magenta).  The upper
right panel shows the sensitivity $S_1$ as a function of $\theta$, for $\phi
=\pi/2$, $N_{\mathrm{tot}} = 2$, and for $\xi = 0.5$ and $r = 0.7$ (red
line), or $\xi = 0.7$ and $r = 0.5$ (blue), or $\xi = 0.5$ and $r = 0.5$
(green). The lower left panel shows 
the ideal sensitivity $S_{1\, \mathrm{min}}$, minimized
with respect to $\beta$, as a function of $\beta_{\mathrm{tot}}$, for
$N_{\mathrm{tot}} = 10^{-1}$ (red line), or $N_{\mathrm{tot}} = 1$
(green), or $N_{\mathrm{tot}} = 10$ (blue), or $N_{\mathrm{tot}} =
10^{2}$ (magenta). The lower right panel shows the optimized 
sensitivity $S_{1\, \mathrm{min}}$ as a function of 
$N_{\mathrm{tot}}$ (solid blue line). Also the shot-noise 
limit $1/ \sqrt{N_{\rm tot}}$ (dashed
red line) is shown.} \label{f4:s1MZ}
\end{figure}

We first address the ideal case, i.e. when the quantum efficiency 
$\eta$ is equal to one (we assume that the two detectors are equal and have 
the same quantum efficiency). Upon employing the 
parameterization introduced in Sect.~\ref{s4.1:qfipi} 
for the input signals and using Eq.~(\ref{sens}), the sensitivity
$S_1$ for the difference photocurrent may be evaluated analytically.
We do not report its full expression since it is 
cumbersome and proceed with the numerical minimization.
As a first step we minimize $S_1$ over the phase shift $\phi$ 
and  the squeezing phase $\theta$: the typical behavior of $S_1$
as a function of these parameters is shown in the upper panels 
of Fig.~\ref{f4:s1MZ} for different possible input 
configurations. As it is apparent from 
the plots $S_1$ is minimized by $\phi = \pi/2$ and $\theta = 0$
independently on the other parameters. We confirmed this by scanning
the full parameter range. 
\par
Using this results we may write a simpler expression for $S_1$ in terms
of the remaining parameters, which reads as follows
\begin{align}\label{S1_MZI}
S_1=&\, \bigg\{ 4\, (1 - \beta) \beta\,  
N_{\mathrm{tot}}^2 + 4\, \sqrt{\beta  N_{\mathrm{tot}}} \sqrt{\beta 
N_{\mathrm{tot}}+1}\nonumber\\
&\times \left((\beta_{\mathrm{tot}}-1) N_{\mathrm{tot}}-\sqrt{N_{\mathrm{tot}} (\beta_{\mathrm{tot}}-\beta )}
\sqrt{N_{\mathrm{tot}} (\beta_{\mathrm{tot}}-
\beta )+1}\right)\nonumber\\
&+ 2\, N_{\mathrm{tot}} \bigg\}^{\frac{1}{2}}/\left( 
\sqrt{2}\, \left| N_{\mathrm{tot}} - 2\, 
N_{\mathrm{tot}} \beta \right| \right)
\end{align}
Results of the minimization of $S_1$ with respect to the squeezing
fractions $\beta_{\mathrm{tot}} \in [0,1]$ and 
$\beta \in [0, \beta_{\mathrm{tot}}]$ are shown in the lower panels 
of Fig.~\ref{f4:s1MZ}. On the left we show $S_1$, numerically minimized 
with respect to $\beta$, as function of $\beta_{\mathrm{tot}}$. We 
find that $S_1$ achieves its minimum for
$\beta_{\mathrm{tot}} = 1$, that is, when all the energy 
is provided by the squeezing or,
equivalently, no coherent (classical) radiation is needed.
Thus, if we fix $\beta_{\mathrm{tot}} = 1$, we can analytically evaluate
the optimal value of $\beta$, which turns out to be $1/2$.
Therefore, the optimal input system is described by two squeezed-vacuum
states, which have the same phase and number of photons.
Overall, the sensitivity $S_{1\, \mathrm{min}} = N_{\mathrm{tot}}^{-1}$,
saturates the Heisenberg limit as a function of the total energy 
$N_{\mathrm{tot}}^{-1}$, see the lower right panel of
Fig.~\ref{f4:s1MZ}.
\subsubsection{Non unit quantum efficiency}
\begin{figure}[h!]
\includegraphics[width=0.49\columnwidth]{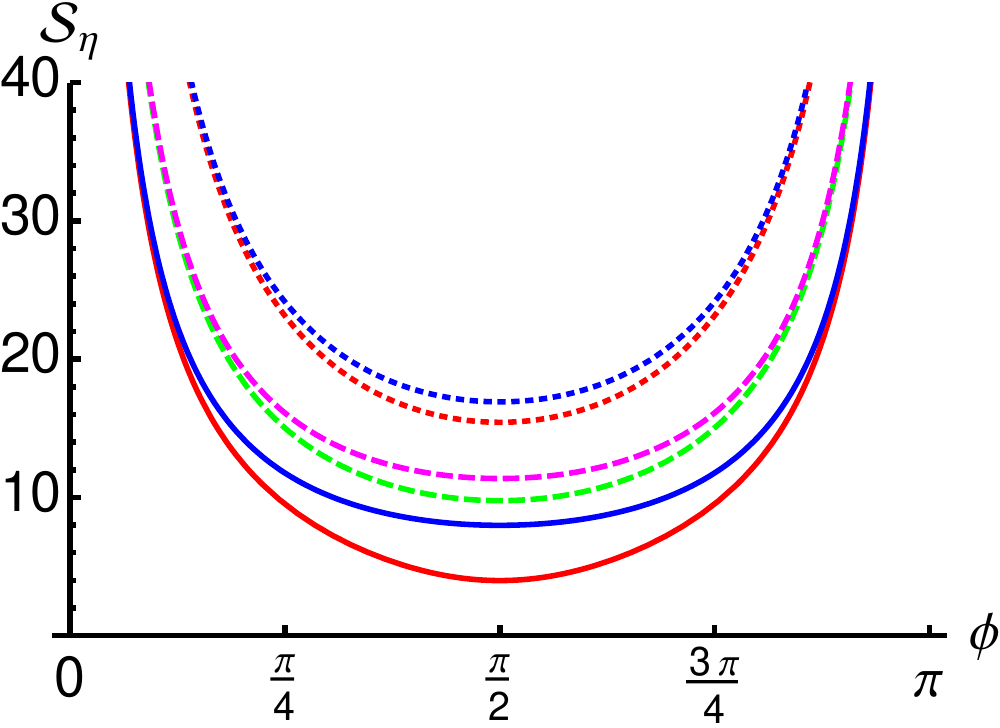}
\includegraphics[width=0.49\columnwidth]{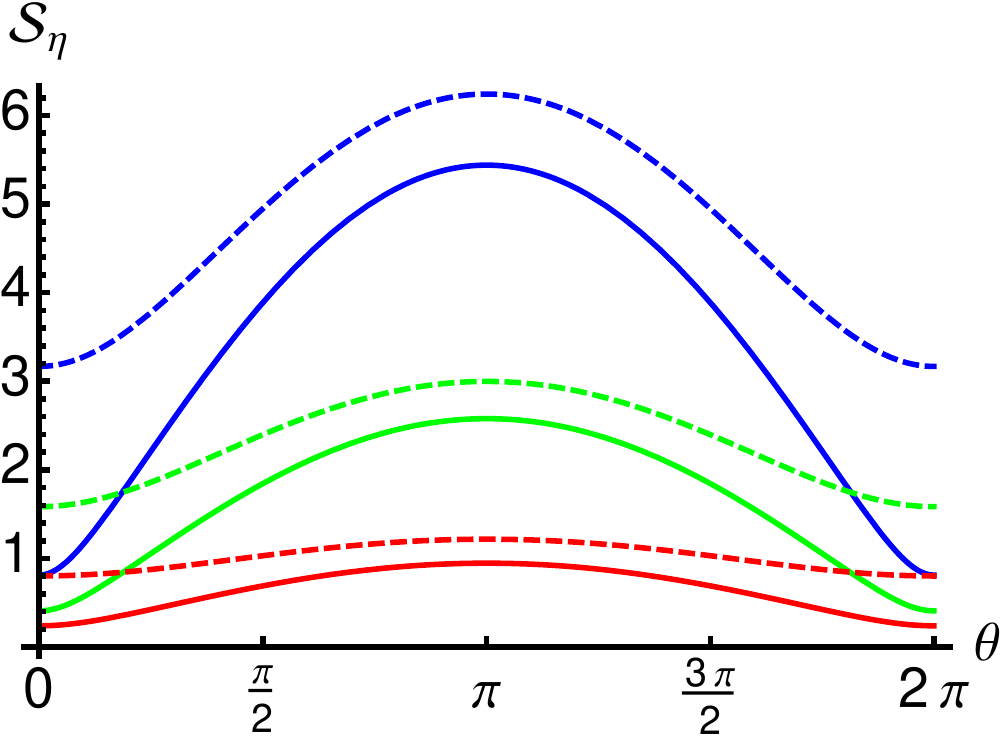}
\includegraphics[width=0.49\columnwidth]{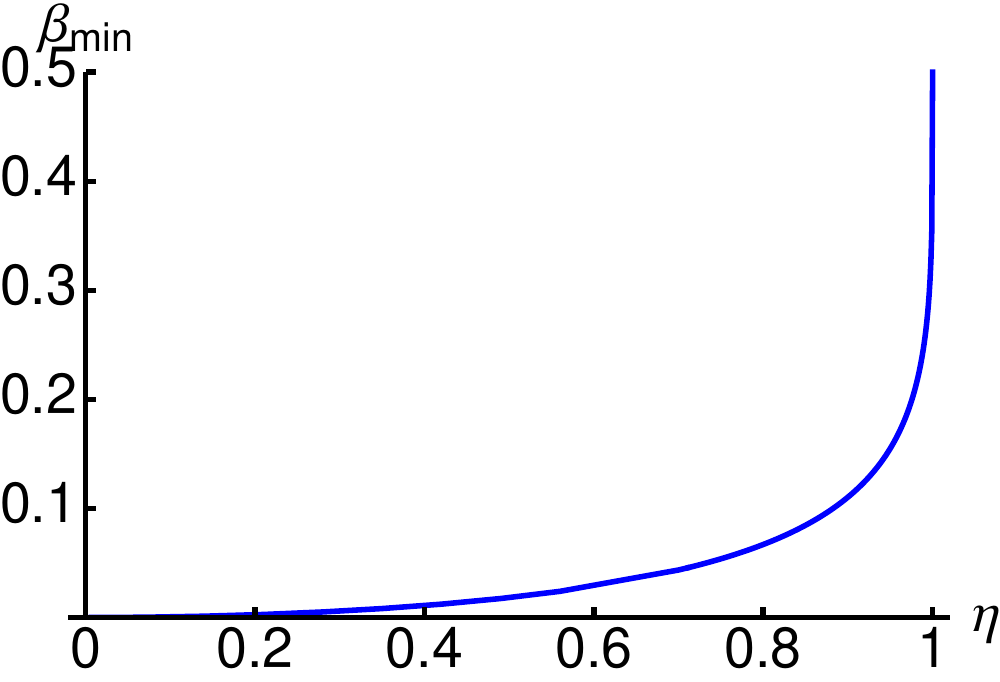}
\includegraphics[width=0.49\columnwidth]{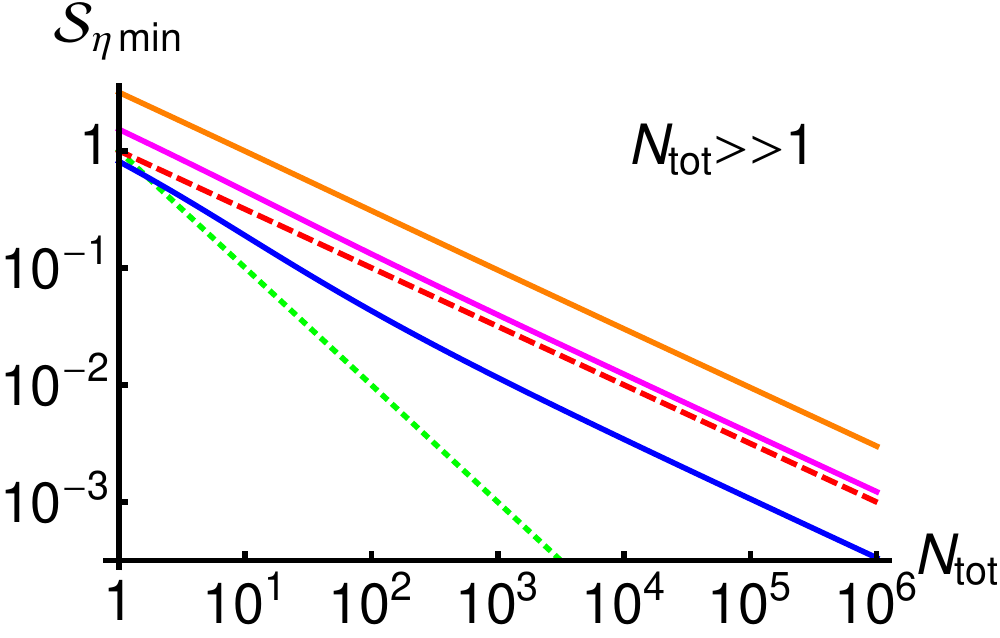}
\caption{(Color Online)
The passive/passive interferometer with realistic phodetection.
In the upper left panel we show the sensitivity $S_{\eta}$ as a 
function of $\phi$, for $N_{\mathrm{tot}} = 1$, $\beta_{\mathrm{tot}} 
= 0.8$, and $\beta = 0.6$. Solid lines
for $\eta = 0.9$, and $\theta = \pi/4$ (red) or $\theta = 3/4\, \pi$ (blue).
The dashed lines for $\eta = 0.3$, with $\theta = \pi/2$ (green)
or $\theta = \pi$ (magenta). The dotted lines for $\eta = 0.1$,
and $\theta = \pi/4$ (red) or $\theta = 3/4\, \pi$ (blue). In the upper
right panel: the sensitivity $S_{\eta}$ as a function of $\theta$, 
for $\phi = \pi/2$, $N_{\mathrm{tot}} = 10$, and 
$\beta_{\mathrm{tot}} = 0.9$. Solid lines for $\eta = 0.8$,
dashed lines $\eta = 0.2$, and for $\beta = 0.1$ (red), 
or $\beta = 0.3$ (green), or $\beta = 0.6$ (blue). In the 
lower right panel we show the fraction 
$\beta_{\mathrm{min}}$, minimizing the sensitivity $S_{\eta}$
for $N_{\rm tot} \ll 1$, as a function of $\eta$. In the lower right
panel: the optimized sensitivity for $N_{\rm tot} \gg 1$ as a function of
$N_{\mathrm{tot}}$, for $\eta = 0.9$ (blue), or $\eta = 
0.4$ (magenta), $\eta = 0.1$ (orange). The shot-noise 
(red dashed line) and the Heisenberg limit (green dotted line) are also 
shown.}
\label{f5:srMZ}
\end{figure}

In the realistic situation the detection efficiency is
lower than one: the sensitivity $S_{\eta}(N_{\mathrm{tot}}, 
\phi, \beta_{\mathrm{tot}}, \beta, \theta)$ depends now on 
all the previous parameters and also on the quantum 
efficiency $\eta$ of the detectors (we assume the same for both).
In order to minimize $S_\eta$ we proceed as in the ideal case.
The behaviour of $S_{\eta}$ as a function of  the phase shift $\phi$ and
the squeezing phase $\theta$ is shown in the upper panels
Fig.~\ref{f5:srMZ}. As we found in the ideal case, 
the sensitivity is minimized for $\phi = \pi/2$ and $\theta = 0$.
\par
In order to solve the optimization problem, we now consider 
the sensitivity $S_\eta$ in the low-energy regime ($N_{\mathrm{tot}} \ll 1$)
and in the high-energy regime ($N_{\mathrm{tot}} \gg 1$).
When $N_{\mathrm{tot}} \ll 1$, the sensitivity can be expanded
to the leading term as follow
\begin{equation}
S_{\eta\, \mathrm{min}}(N_{\mathrm{tot}},\beta_{\mathrm{tot}},\beta) \approx 
\sqrt{
\frac{1 - 2\eta \sqrt{\beta (\beta_{\mathrm{tot}} - \beta)}}
{\eta  (1 - 2 \beta)^2 }}
\, \frac{1}{\sqrt{N_{\mathrm{tot}}}}.
\end{equation}
The coefficient multiplying $1/\sqrt{N_{\mathrm{tot}}}$ is minimised
when $\beta_{\mathrm{tot}} = 1$, and $\beta$ depends on $\eta$ as
shown in the lower left panel of Fig.~\ref{f5:srMZ}. 
For $\eta < 1$, we find that the optimal working regime 
is obtained when we introduce more squeezed photons
in one arm than in the other.
The optimized sensitivity is then given by
\begin{equation}
S_{\eta\, \mathrm{min}}(N_{\mathrm{tot}}) =
\sqrt{ \frac{1}{2 \eta} + \frac{1}{2}\, \sqrt{\frac{1 - \eta^2}{\eta^2}}}\,
\frac{1}{\sqrt{N_{\mathrm{tot}}}}.
\end{equation}
\par
Let us now focus to the high-energy regime ($N_{\mathrm{tot}} \gg 1$),
in which the sensitivity can be expanded as
\begin{equation}\label{Se_her}
S_{\eta\, \mathrm{min}}(N_{\mathrm{tot}},\beta) \approx 
\sqrt{\frac{1-\eta}{\eta (1 - 2 \beta)^2}}\,
\frac{1}{\sqrt{N_{\mathrm{tot}}}},
\end{equation}
where the coefficient $\beta_{\mathrm{tot}}$ does not explicitly appear,
but affects the coefficients of higher orders. However, we can safety
set $\beta_{\mathrm{tot}} \neq 1$ and consider only the first order.  In
order to minimise the sensitivity of Eq.~(\ref{Se_her}), we can set
$\beta = 0$ (or $\beta=1$), that is, all the squeezing photons can be
injected in one arm. Thus, the optimal sensitivity turns
out to be 
\begin{equation}
S_{\eta\, \mathrm{min}}(N_{\mathrm{tot}}) = 
\sqrt{\frac{1 - \eta}{\eta}}\, \frac{1}{\sqrt{N_{\mathrm{tot}}}}
\end{equation}
and we show its behaviour, for different values of $\eta$, in 
the lower right panel of Fig.~\ref{f5:srMZ}. 
\par
It is worth noting that the sensitivity in presence of noise
scales as the shot noise limit, that is $1/\sqrt{N_{\mathrm{tot}}}$. 
Therefore, using quantum (squeezed) radiation inside an interferometer
allows to improve the sole coefficient multiplying the shot-noise 
limit \cite{oli:par:OptSp}.
\subsection{The passive/active case} \label{s5:piam}
This interferometer is composed by a beam splitter, mixing the light
beams before the phase shift, and by an active measurement scheme
where an amplifier is used to recombine the modes, adding
squeezing before the photodetection. 
The input states for this interferometers are
$\ket{\alpha, \xi}\ket{\gamma}$, where $\alpha, \gamma \in \R$ are the
coherent amplitudes and $\xi = r\, e^{-i \theta}$ is the squeezing
coefficient of the first state, $r \in \R^+$ and $\theta \in [0,2\,
\pi)$.  Furthermore, the beam splitters is balanced and its phase is set
equal to zero, while the OPA is parametrised by a squeezing coefficient
$\zeta = r_1\, e^{-i \theta_1} \in \C$, where $r_1 \in \R^+$, and
$\theta_1 \in [0, 2 \pi)$.
\subsubsection{Ideal photodetection}
\begin{figure}[tb]
\includegraphics[width=0.49\columnwidth]{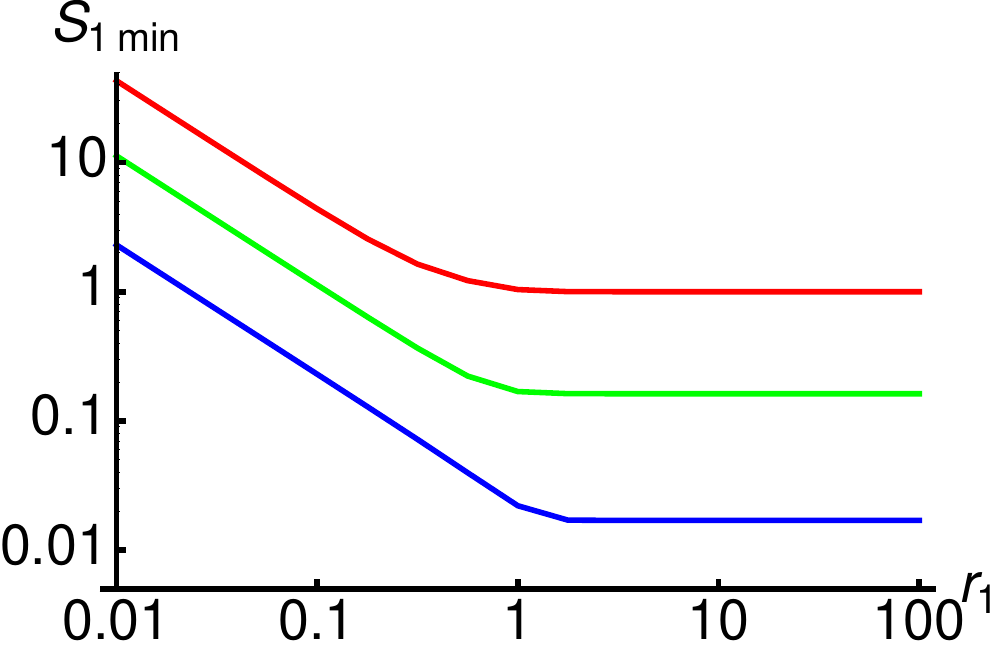}
\includegraphics[width=0.49\columnwidth]{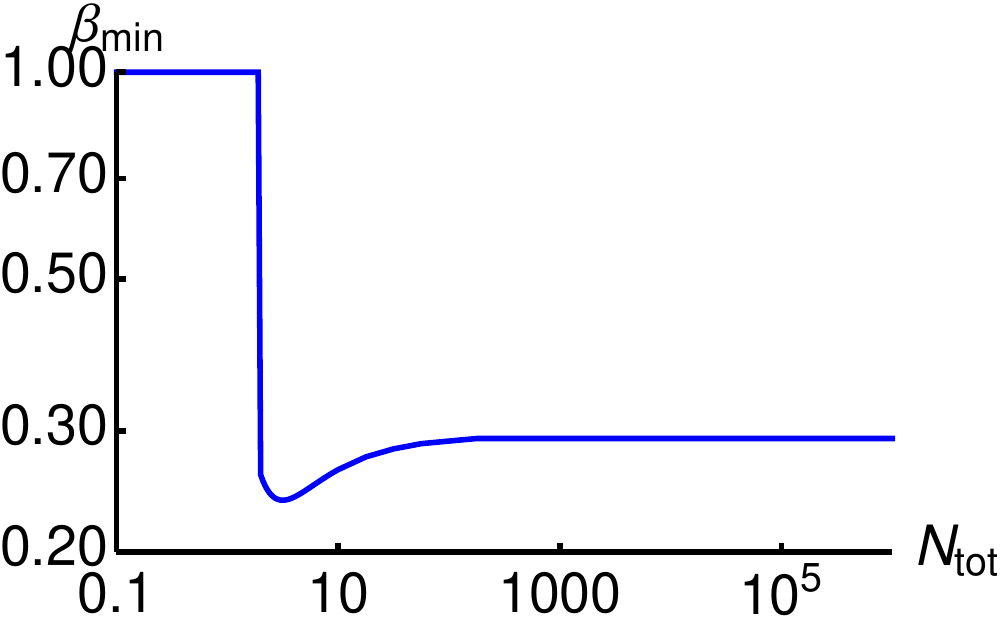}
\includegraphics[width=0.49\columnwidth]{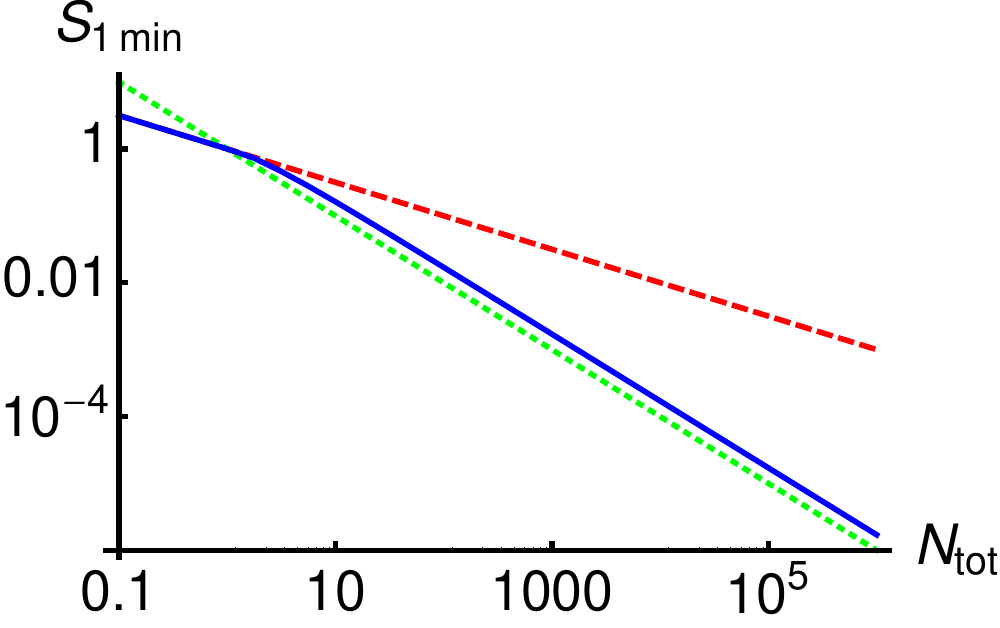}
\includegraphics[width=0.49\columnwidth]{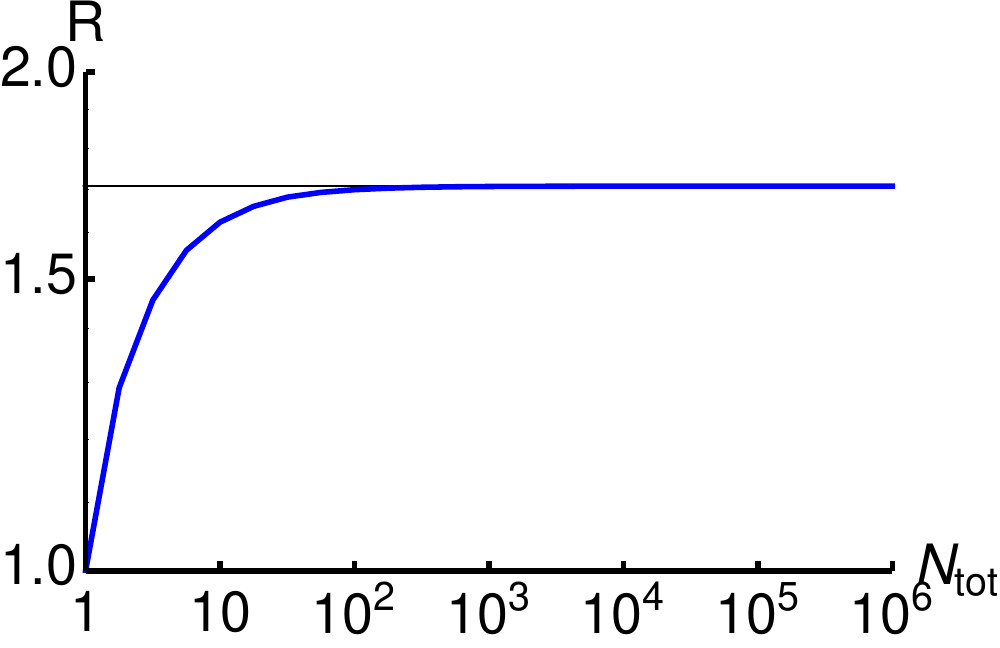}
\caption{(Color online) The passive/active interferometer with ideal
photodetection. The upper left panel shows the minimized sensitivity 
$S_{1\, \mathrm{min}}$ as a function of the parameter $r_1$, for 
$N_{\mathrm{tot}} = 1$ (red line), $N_{\mathrm{tot}} = 10$ 
(green line), and $N_{\mathrm{tot}} = 100$ (blue line). 
The upper right panel shows the parameter $\beta$ minimizing 
the sensitivity as a function of $N_{\mathrm{tot}}$.
The lower left panel shows the optimal sensitivity 
$S_{1\, \mathrm{min}}$ as a function of
the total number of photons $N_{\mathrm{tot}}$, 
together with the shot-noise limit
(red dashed line) and the Heisenberg limit (green dotted line).
The lower right panel illustrate the behaviour of the ratio 
$R = S_{1\, \mathrm{min}}/S_{\mathrm{HL}}$, where
$S_{\mathrm{HL}} = 1/N_{\mathrm{tot}}$ is the Heisenberg 
scaling, as a function of $N_{\mathrm{tot}}$. 
The asymptotic value of $R$ is $1 + 1/\sqrt{2}$.
}
\label{f6:pasact}
\end{figure}
\par
The ideal $S_1$ depends on the coherent amplitudes 
$\alpha$ and $\gamma$ and on the squeezing ones $\xi = r e^{-i \theta}$,
and $\zeta = r_1 e^{-i \theta_1}$.  In the following we will
use the same parameterization introduced in sect.~\ref{s4.1:qfipi}.
The ideal sensitivity is a cumbersome function
$S_1(N_{\mathrm{tot}}, \delta, \phi,\beta, \theta, r_1, \theta_1)$,
which we do not report here. However, from its analytic expression, one
may observe that the phase of the OPA, $\theta_1$, can be set equal to
zero without loss of generality. We now focus on the behavior of the
sensitivity when the amplifier introduces squeezed photons in the
system. As we have done in the previous cases, we have numerically 
minimized $S_1$ with respect to
all the parameters except $N_{\rm tot}$ and $r_1$.  The behaviour  
of sensitivity is shown in the upper left panel of Fig.~\ref{f6:pasact}: 
in the limit $r_1 \gg 1$, the sensitivity reaches its minimum.  
Therefore, the optimal sensitivity is obtained when a large number 
of squeezed photons is introduced by the amplifier.  
\par
From the analysis of the numerical minimization, we find that the
optimal value of the coherent trade-off coefficient $\delta$ is 1. In
this case, the best configuration for the estimation of the phase is
obtained when the input system is described by the state
$|\Psi_\ip\rangle\rangle=\ket{\alpha,\xi} \ket{0}$.  Concerning 
the squeezing fraction $\beta$, we found that the optimal 
value $\beta_{\hbox{min}}$ minimising $S_1$ depends in a non-trivial 
way on the number of photons
of the input state, as shown in the upper right panel of
Fig.~\ref{f6:pasact}.
Remarkably, in the limit of high energy the parameter is
$\beta_{\mathrm{min}} \approx 0.3$. Thus, the optimal estimation of
$\phi$ is obtained when both coherent and squeezed photons are used.  In
the low-energy range, the value of $\beta_{\mathrm{min}}$ is equal to 1,
and thus the optimal state of the system is $\ket{0,\xi}\ket{0}$: only
squeezed light is needed.  
\par
The optimized  sensitivity $S_{1\, \mathrm{min}}$ is shown in
the the lower left panel of Fig.~\ref{f6:pasact}.  
In the high-energy regime, the
sensitivity is proportional to the Heisenberg limit, even if 
with coefficient larger than one. 
In the low-energy range, the sensitivity goes down to the
shot-noise limit.  In order to assess the sensitivity
when $N_{\mathrm{tot}} \gg 1$, we consider the ratio $R = S_{1\, 
\mathrm{min}}/S_{\mathrm{HL}}$ between the numerically minimized 
$S_{1\, \mathrm{min}}$ and the Heisenberg scaling (Fig.~\ref{f6:pasact}, 
lower right panel): we find that, in the high-energy
regime
\begin{equation}
R \stackrel{N_{\mathrm{tot}}\gg 1}{\longrightarrow}
\left(1 + \frac{1}{\sqrt{2}}\right)\,. 
\end{equation}
\subsubsection{Non unit quantum efficiency}
\begin{figure}[h!]
\center
\includegraphics[width=0.9\columnwidth]{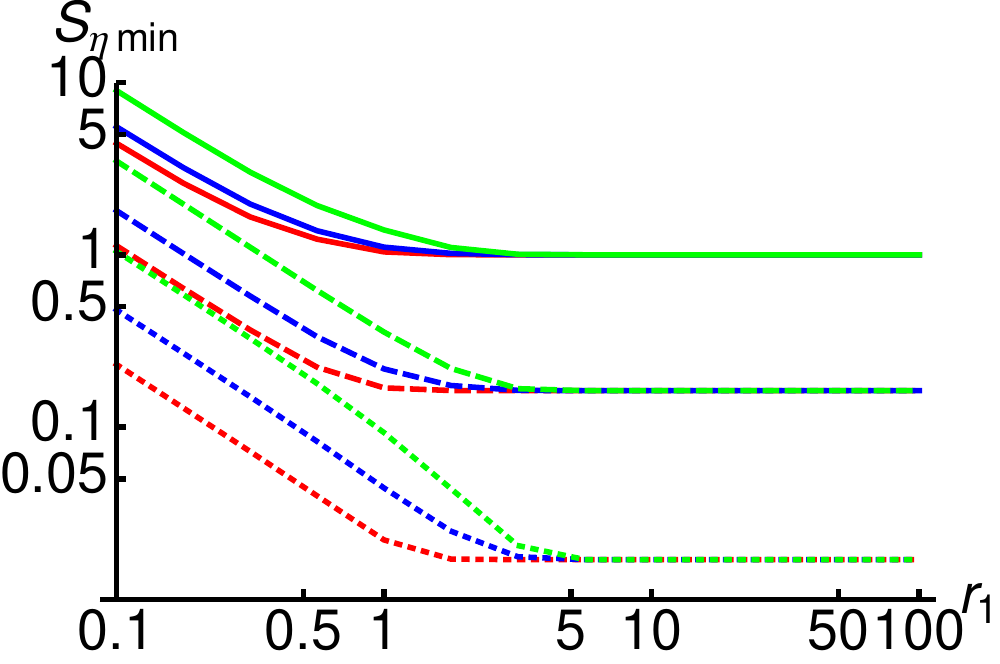}
\caption{(Color online) The passive/active interferometer with 
realistic photodetection. The plot shows the optmized sensitivity 
$S_{\eta\, \mathrm{min}}$ as a function of $r_1$, for 
$N_{\mathrm{tot}} = 1$ (solid line), $N_{\mathrm{tot}} = 10$ 
(dashed line), and $N_{\mathrm{tot}}
= 100$ (dotted line), and for $\eta = 1$ (red line),
$\eta = 0.6$ (blue line), and  $\eta = 0.2$ (green line).}
\label{CLBI3s_Se}
\end{figure}

For non unit quantum efficiency the sensitivity is
expressed as a function $S_{\eta}(N_{\mathrm{tot}}, \delta, \phi, \beta,
\theta, r_1, \theta_1)$.  We analyze its behaviour starting from
its dependence on the parameter $r_1$, which corresponds to the 
energy (average number of squeezed photons) introduced by the 
parametric amplifier.  In the ideal case,
we have seen that the optimal sensitivity is obtained for $r_1 \gg
1$.  To analyze the behavior of $S_1$ in the presence of 
detection loss, we minimize it with respect to all the 
parameters, except for
$N_{\mathrm{tot}}$, $\eta$, and $r_1$. The optimal value 
$S_{\eta}(N_{\mathrm{tot}},r_1)$ is shown in Fig.~\ref{CLBI3s_Se}, for
different values of the total number of photons $N_{\mathrm{tot}}$ and
quantum efficiency $\eta$.  From the plot we can extract two main
results. First of all, also for non unit quantum efficiency, 
the sensitivity is optimized for $r_1 \gg 1$.  Therefore, in the optimal
configuration, we have to provide as many photons as possible through
the amplifier.  Moreover, if a large amount of energy is introduced
inside the system with the amplifier, sensitivity approaches the
ideal one, as it may be seen by taking the limit of $S_{\eta}$
for $r_1 \rightarrow \infty$.  Therefore, the active measurement stage
allows us to balance the losses of the detectors, and to obtain the
ideal sensitivity even in the presence of noise.  Since the real
sensitivity is minimized for $r_1 \gg 1$, and we have shown that in that
regime the real and ideal sensitivities coincide, we can use the same
results from the ideal case. Thus, we find that the sensitivity is
still proportional to the Heisenberg limit, even if the system is
affected by noise.
\subsection{The active/passive case}
Now we turn our attention to the performances of the active
interferometer of Sec.~\ref{s4.2:qfiai}, when a passive measurement
stage is employed.  The input beams are described by two coherent
states, $\ket{\alpha}$ and $\ket{\gamma}$, respectively, where we take
$\alpha, \gamma \in {\mathbbm R}$.  The first component of the
interferometer is the OPA, described by the operator $U_{\OPA}(\xi)$,
with $\xi = r e^{- i \theta}$, where $r \in \R^+$, and $\theta \in [0, 2
\pi)$.  Finally, the beam splitter inside the measurement stage is
assumed to be balanced.
\subsubsection{Ideal photodetection}
In order to analyze the sensitivity $S_1$ in the ideal scenario, i.e.
$\eta = 1$, we consider the following parameters: First, the total
number of photons, here given by $N_{\mathrm{tot}} = \left( \alpha^2 +
\gamma^2 + 1 \right) \cosh 2 r + 4\, \alpha \gamma \cos \theta \sinh r
\cosh r - 1$, which accounts for both the photons in the input signals
as well as those introduced by the amplifier. Then, we consider the
ratio $\delta$ between the number of coherent photons in the two input
states, as defined in Sec.~\ref{s4.1:qfipi}, and the squeezing fraction
$\beta = 2\, \sinh^2 r/N_{\mathrm{tot}}$, expressing the ratio between
the number of squeezed photon injected by the amplifier and the total
number of photons. In the following, we thus consider $S_1
= S_1(N_{\mathrm{tot}},\delta,\phi,\beta,\theta)$.
\par
As in the previous sections, we want to minimize $S_1$ to obtain a
function of the sole average number of photons $N_{\mathrm{tot}}$. To
this aim, we numerically minimize $S_1$ and find that the coherent
trade-off coefficient $\delta$ should be $\delta=1/2$ (the procedure is 
similar to previous cases, we do not report the details), i.e. the 
number of coherent photons in the input signals should be balanced.  
Similarly, one finds that the optimal value 
$\phi=\pi/2$ is independent on the other parameters.
\begin{figure}[h!]
\includegraphics[width=0.49\columnwidth]{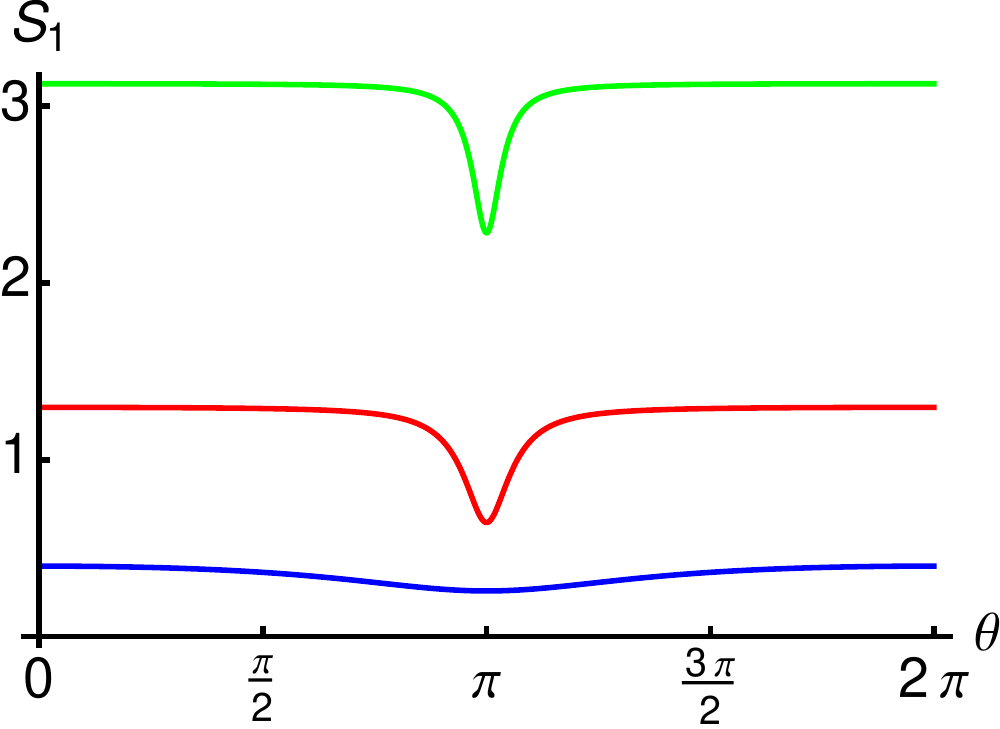}
\includegraphics[width=0.49\columnwidth]{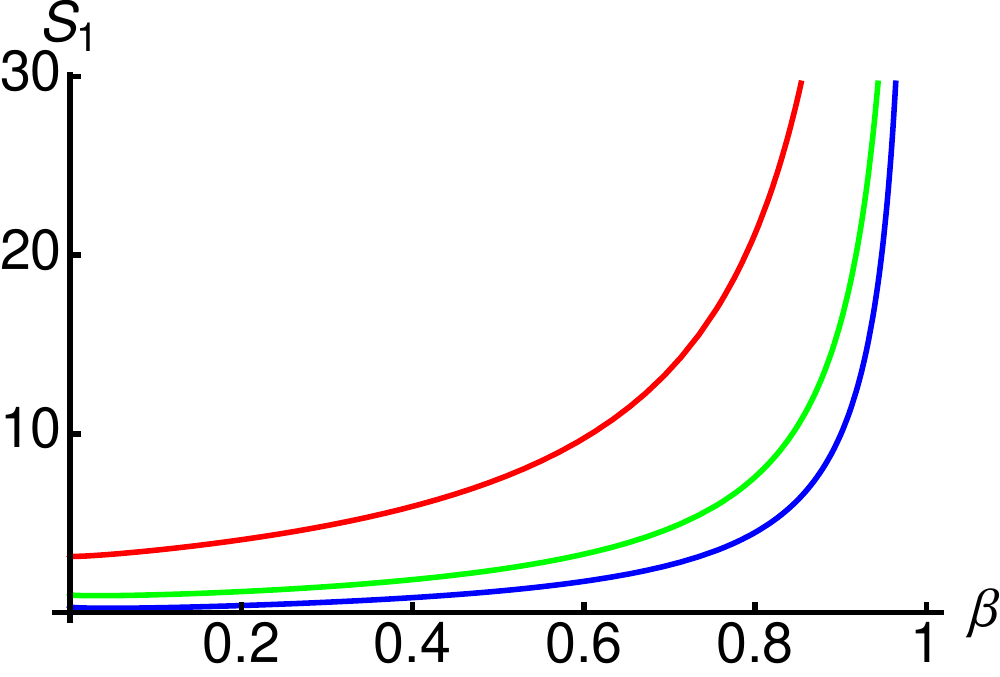}
\includegraphics[width=0.49\columnwidth]{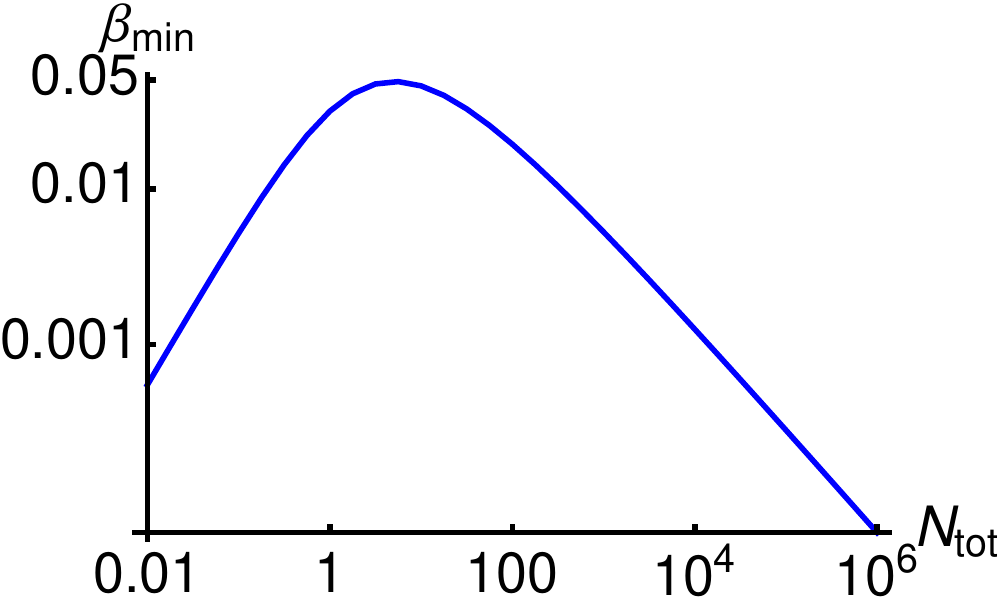}
\includegraphics[width=0.49\columnwidth]{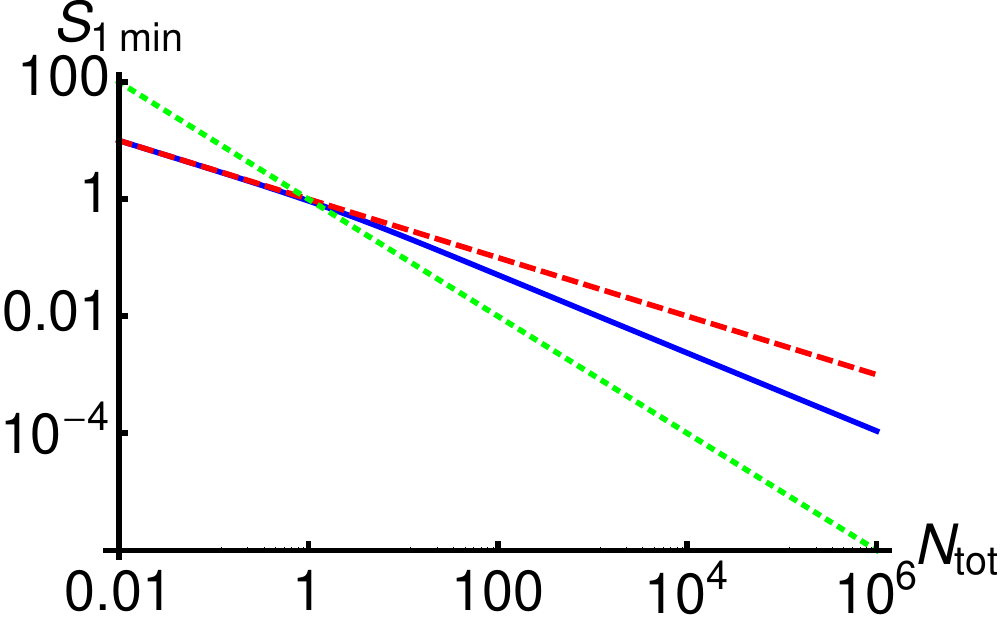}
\caption{(Color online) 
The active/passive interferometer with ideal photodetection. In the
upper left panel we show the sensitivity $S_1$ as a function of 
$\theta$, when
$N_{\mathrm{tot}} = 10$, $\delta = 1/2$ and $\phi = \pi/2$.
The blue line refers to $\beta = 0.01$, the red line to $\beta = 1/3$,
and the green one to $\beta = 2/3$. The upper right panel shows the 
sensitivity $S_1$ as a function of $\beta$, when
$\delta = 1/2$, $\phi = \pi/2$, and $\theta = \pi$. The red line is taken for
$N_{\mathrm{tot}} = 0.1$, the green one for $N_{\mathrm{tot}} = 1$,
and the blue one for $N_{\mathrm{tot}} = 10$.
The lower left panel shows the value of 
$\beta$ minimizing $S_{1\, \mathrm{min}}$,
as a function of $N_{\mathrm{tot}}$. Finally, the lower right panel
shows the optimized $S_1$ as a function
of $N_{\mathrm{tot}}$ (blue line), with the shot-noise limit (red dashed line)
and the Heisenberg limit (green dotted line).}
\label{f8:actpas}
\end{figure}
\par
We now consider the parameter $\theta$, that is the phase introduced by
the amplifier.  The value of $S_1$ as a function of $\theta$ is shown in
the upper left panel of Fig.~\ref{f8:actpas}, for different values of
the parameter $\beta$. The plot shows that the value of $\theta$
minimizing the sensitivity is $\pi$.  It is worth noting, moreover, that
$S_1$ grows with $\beta$.  Therefore, it seems that the best
estimation possible is achieved for small values of $\beta$.  In the
upper right panel of Fig.~\ref{f8:actpas}, we show the sensitivity as
a function of the parameter $\beta$, showing that $S_1$ is indeed minimized
for values of $\beta$ close to zero. However, a small fraction of
squeezing is necessary, otherwise the sensitivity would be equal to the
shot-noise limit $1/\sqrt{N_{\mathrm{tot}}}$. In the lower left panel
of the same figure  the value  $\beta_{\mathrm{min}}$ 
minimizing $S_1$ is shown as a function of
$N_{\mathrm{tot}}$. It grows with $N_{\mathrm{tot}}$ until it reaches a
maximum for $N_{\mathrm{tot}} \approx 10$, close to the value 
$\beta_{\mathrm{min}}\simeq 0.05$, and
then it decreases as the total number of photons increase.  
\par
After the minimization of $S_1$, it is interesting
to analyze how it behaves as the total number $N_{\rm tot}$ of photons is
changed. This is illustrated in the lower right panel of
Fig.~\ref{f8:actpas}.
In the low-energy regime ($N_{\mathrm{tot}} \ll 1$), the optimal 
sensitivity is equal to the shot noise limit, whereas in the 
high-energy regime, $N_{\mathrm{tot}} \gg 1$), we have that $S_{1\, 
\mathrm{min}}$ is below the
the shot-noise limit, but above the Heisenberg limit.
Actually, $S_{1\, \mathrm{min}}$ is proportional to 
$N_{\mathrm{tot}}^{-2/3}$
in the high-energy regime.
\subsubsection{Non unit quantum efficiency}
In a realistic situation, when $\eta < 1$, the sensitivity is a
function $S_{\eta}(N_{\mathrm{tot}}, \delta, \beta, \phi, \theta)$.  An
analytic, though cumbersome, expression for $S_{\eta}$ may 
obtained: we are not reporting it here. We start by noting 
that, as it happens for the other configurations, the 
sensitivity $S_{\eta}$ is still minimized by $\delta = 1/2$,
$\phi=\pi/2$, i.e. also for the active/passive case a non unit 
quantum quantum efficiency does not influence the
position of the minimum of the sensitivity, at least
for what concerns the
phases of the system.
\begin{figure}[h!]
\includegraphics[width=0.49\columnwidth]{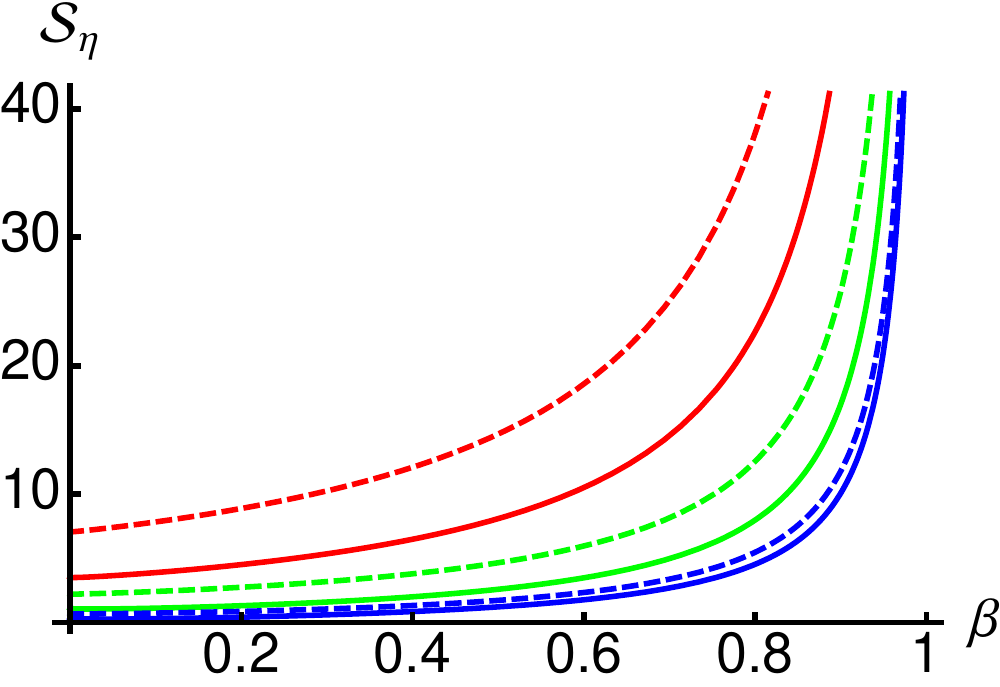}
\includegraphics[width=0.49\columnwidth]{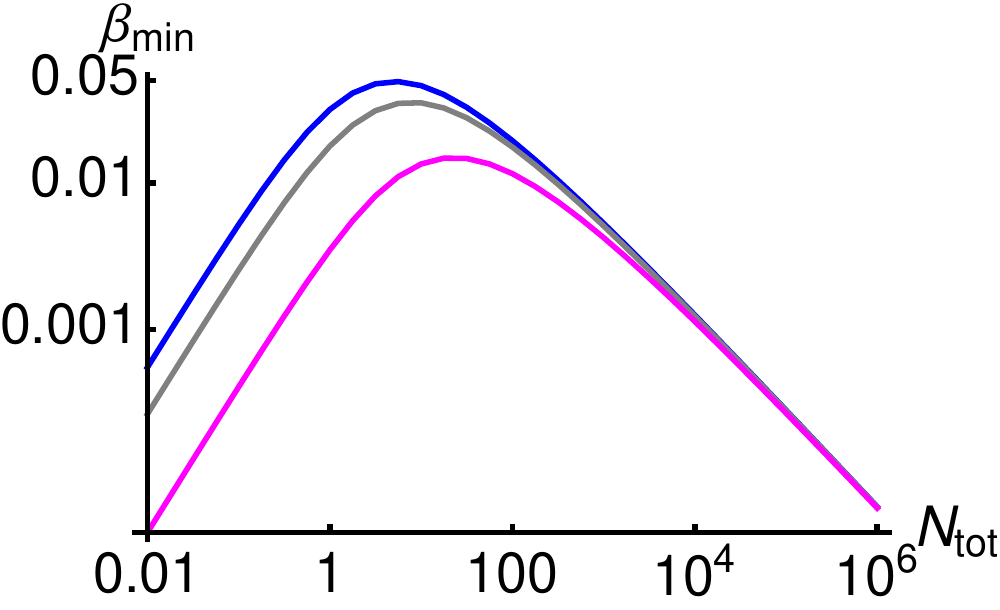}
\includegraphics[width=0.49\columnwidth]{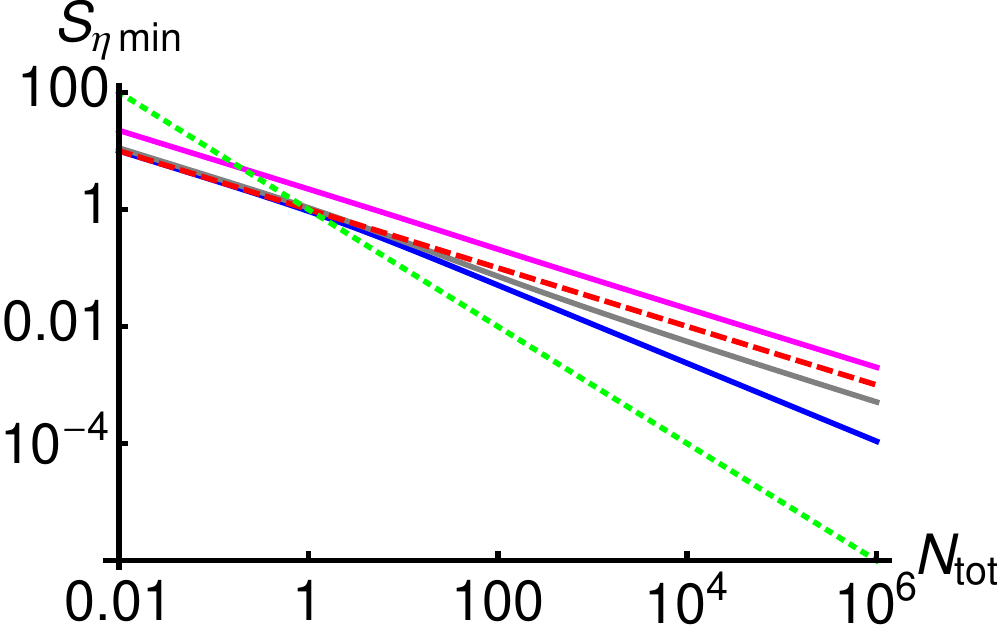}
\includegraphics[width=0.49\columnwidth]{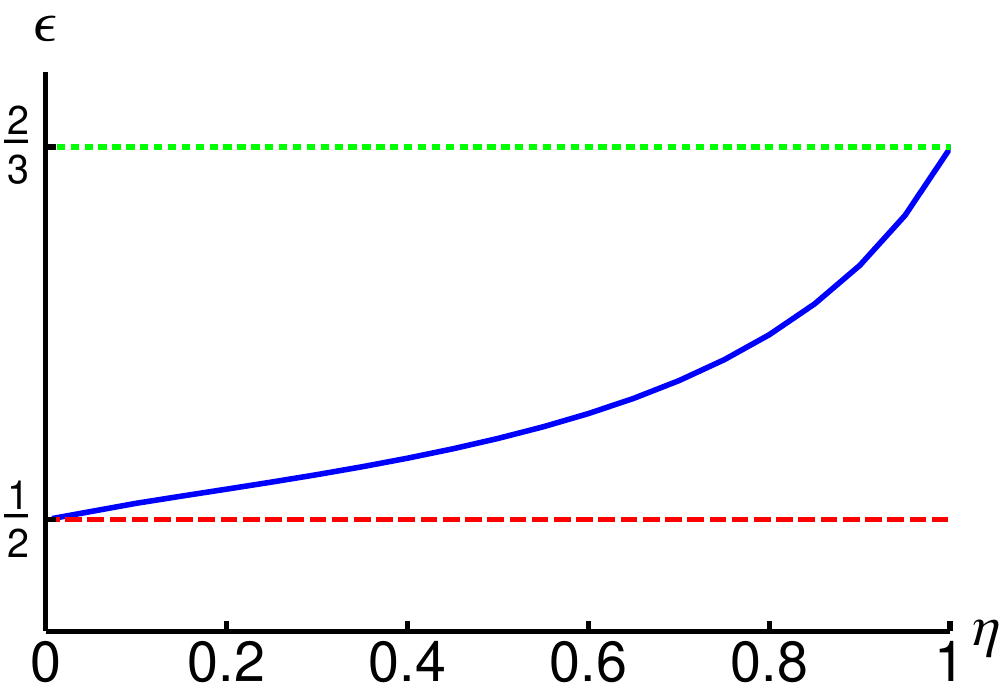}
\caption{(Color online) The active/passive inteferometer with realistic
photodetection. The upper left panel shows the sensitivity $S_{\eta}$ as a
function of $\beta$, for $\delta = 1/2$, $\phi = \pi/2$, and $\theta =
\pi$. The solid lines are for $\eta = 0.8$, and the dashed lines for
$\eta = 0.2$. The red lines are for $N_{\mathrm{tot}} = 0.1$, green
lines for $N_{\mathrm{tot}} = 1$, and blue lines for $N_{\mathrm{tot}} =
10$. In the upper right panel the value of $\beta$ minimizing 
$S_{\eta}$ is shown as a function of $N_{\mathrm{tot}}$.
The blue line refers to $\eta = 1$, the grey one to $\eta = 0.6$,
and magenta one to $\eta = 0.2$. The lower left panel shows the 
optimized sensitivity $S_{\eta\, \mathrm{min}}$
as a function of $N_{\mathrm{tot}}$, for $\eta = 1$ (blue line), $\eta = 0.8$
(grey line) and $\eta = 0.2$ (magenta line). The shot-noise limit (red dashed
line) and the Heisenberg limit (green dotted line) are shown. The 
lower right panel shows the exponent $\epsilon(\eta)$ as a function of 
the quantum efficiency.}
\label{f9:actpas}
\end{figure}
\par
Using this result we may write the sensitivity as 
\begin{align}
S_{\eta
}=&\, \Bigg\{\Big[\eta (\beta - 2) \beta ^2  N_{\mathrm{tot}}^2
- \eta \beta + \beta  N_{\mathrm{tot}} (\eta(\beta -3) - 1 )\nonumber\\
&+\sqrt{\beta  N_{\mathrm{tot}} (\beta  
N_{\mathrm{tot}}+2)} (\eta  (1-(\beta -2) \beta
N_{\mathrm{tot}})+1)-1\Big]\nonumber\\
&/\Big[\eta (\beta -1)^2 N_{\mathrm{tot}} \left( 
\sqrt{\beta N_{\mathrm{tot}}
(\beta  N_{\mathrm{tot}} + 2 )} - 1 - \beta 
N_{\mathrm{tot}} \right) \Big] \Bigg\}^{\frac{1}{2}}
\,,
\label{CLBI2_Semin}
\end{align}
which now depends on the total number of photons,
on the squeezing fraction and on the quantum efficiency $\eta$.  We
minimize the sensitivity with respect to the squeezing fraction $\beta$.
In Fig.~\ref{f9:actpas} (upper left panel), the sensitivity is shown as a
function of $\beta$. The best sensitivity is achieved, again, for values
of $\beta$ close to $0$. The actual value of $\beta$ minimizing the
sensitivity is shown in the upper right panel of Fig.~\ref{f9:actpas} 
for different values of $\eta$.  The overall behavior is similar to the
ideal case, even if we need less
and less squeezing as the quantum efficiency decreases.
\par
The optimum sensitivity is thus obtained when two identical
coherent states are injected in the interferometer, and when the
amplifier introduces only a small fraction of squeezed photons. In
the lower left panel of Fig.~\ref{f9:actpas}, the optimized 
sensitivity, as a function
of the total number of photons is plotted for different values of
$\eta$.  Notice that, in the high-energy regime, the sensitivity
gradually approaches the shot-noise limit as the quantum
efficiency decreases.  In the limit $N_{\rm tot}\gg 1$ the 
sensitivity depends on the total number as a power-law 
$N_{\rm tot}^{-\epsilon(\eta)}$, where the function 
$\epsilon(\eta)$ is shown in the lower right panel of
Fig.~\ref{f9:actpas}.  The coefficient $\epsilon$ decreases from
the value $2/3$ to $1/2$ (the shot-noise limit) as the quantum
efficiency decreases.  Compared to the Mach-Zehnder
interferometer studied in sect.~\ref{ss:PassIntPassMeas}, the active
interferometer we have analyzed here has a sensitivity which approaches
the shot-noise limit only for $\eta \rightarrow 0$. Thus, despite this
interferometer does not allow us to reach the Heisenberg limit, 
its sensitivity improves over that of a Mach-Zender
interferometer, at least for $\eta$ not to far from unit.
\subsection{The active/active case}
We now study the sensitivity of the active interferometer of
Sect.~\ref{s4.2:qfiai}, with an active measurement stage.  The
sensitivity of this interferometer was studied in Ref.~\cite{agar:10}, 
though only in the ideal case.  
\par 
The input signals here are two coherent states, $\ket{\alpha}$ in 
the first mode, and $\ket{\gamma}$ in the second one. We can take, 
without loss of generality, both $\alpha$ and $\gamma$ real. 
The interferometer involves two amplifiers, described by the same 
unitary operator $U(\xi_j)$, with $j = 1, 2$. The coefficient  
$\xi_1$ is complex, while $\xi_2$ can takes real values without loss 
of generality. We rewrite $\xi_1 = r_1 e^{-i
\theta_1}$, and $\xi_2 = r_2$, where $r_1, r_2 \in \R^+$, and $\theta_1
\in [0, 2 \pi)$.
\subsubsection{Ideal photodetection}
Te sensitivity is a function $S_1(\alpha, \gamma,
\phi, r_1, \theta_1, r_2)$. As we have done for other configurations, 
we introduce a convenient parametrization to better capture the 
energy lanscape of the system. The first parameter is the total 
number of photons, $N_{\mathrm{tot}}$,
resulting from the input signals and the first amplifier.  This
parameter can be expressed as $N_{\mathrm{tot}} = \left( \alpha^2 +
\gamma^2 + 1 \right) \cosh 2 r_1 + 4\, \alpha \gamma \cos \theta \sinh
r_1 \cosh r_1 - 1$.  The second parameter is the coherent trade-off
coefficient $\delta$, introduced in sect.~\ref{s4.1:qfipi}.  The last
parameter we consider is the squeezing fraction of photons introduced by
the first amplifier, namely, $\beta = 2\, \sinh^2 r_1/N_{\mathrm{tot}}$,
taking values between $0$ and $1$.  The ideal sensitivity turns out to
be a function $S_1(N_{\mathrm{tot}}, \delta, \phi, \beta, \theta_1,
r_2)$.
\par
\begin{figure}[h!]
\includegraphics[width=0.49\columnwidth]{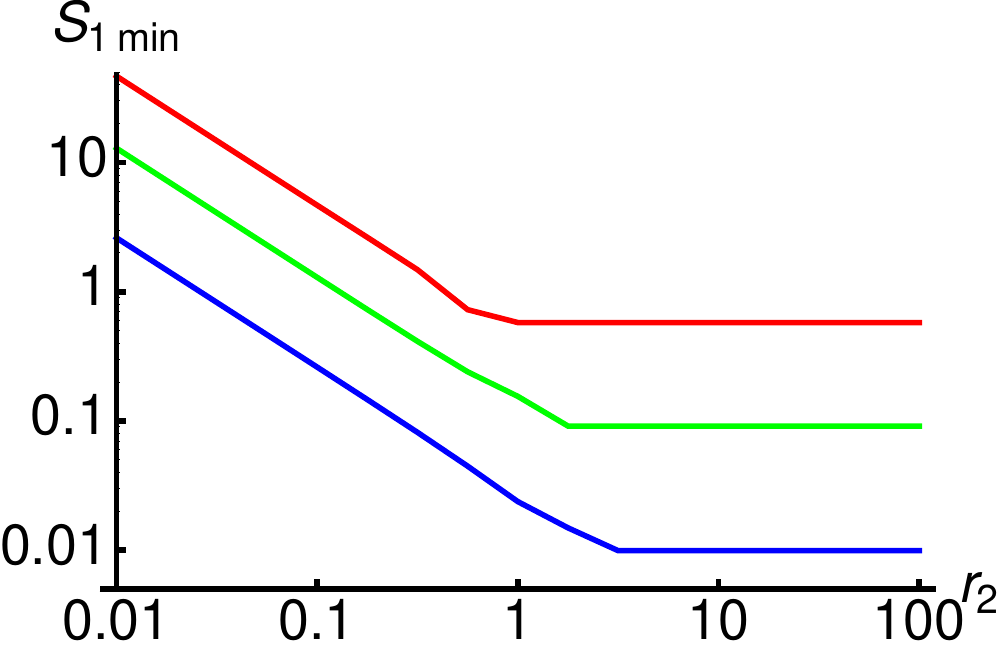}
\includegraphics[width=0.49\columnwidth]{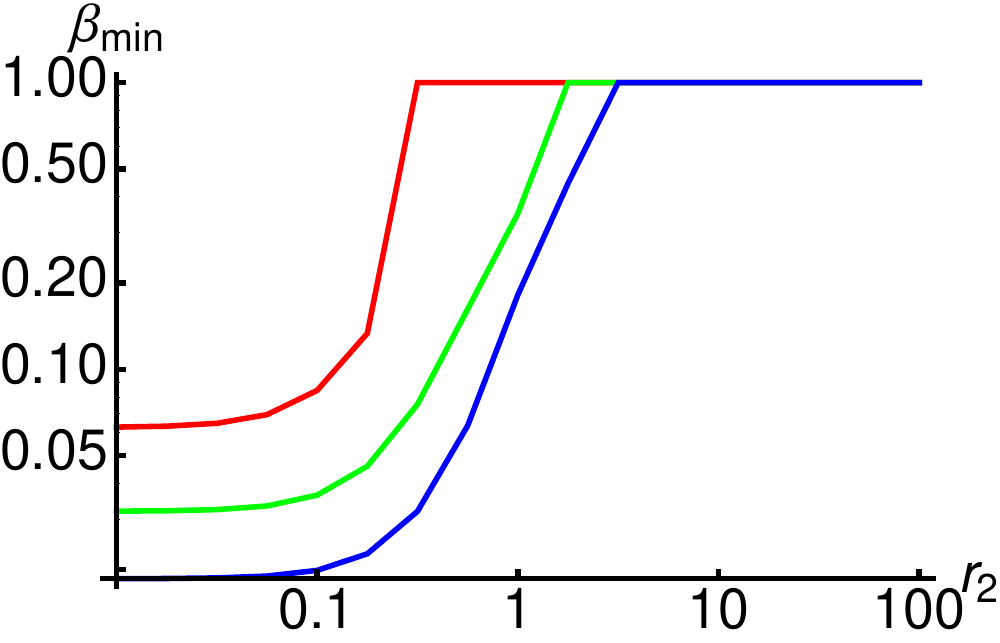}
\includegraphics[width=0.49\columnwidth]{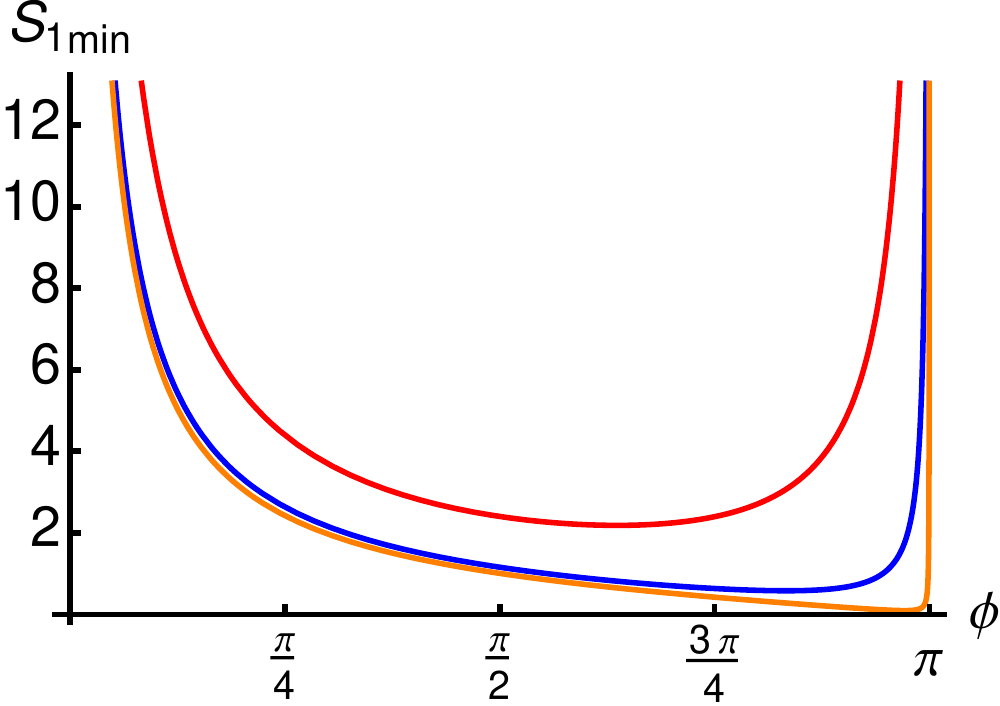}
\includegraphics[width=0.49\columnwidth]{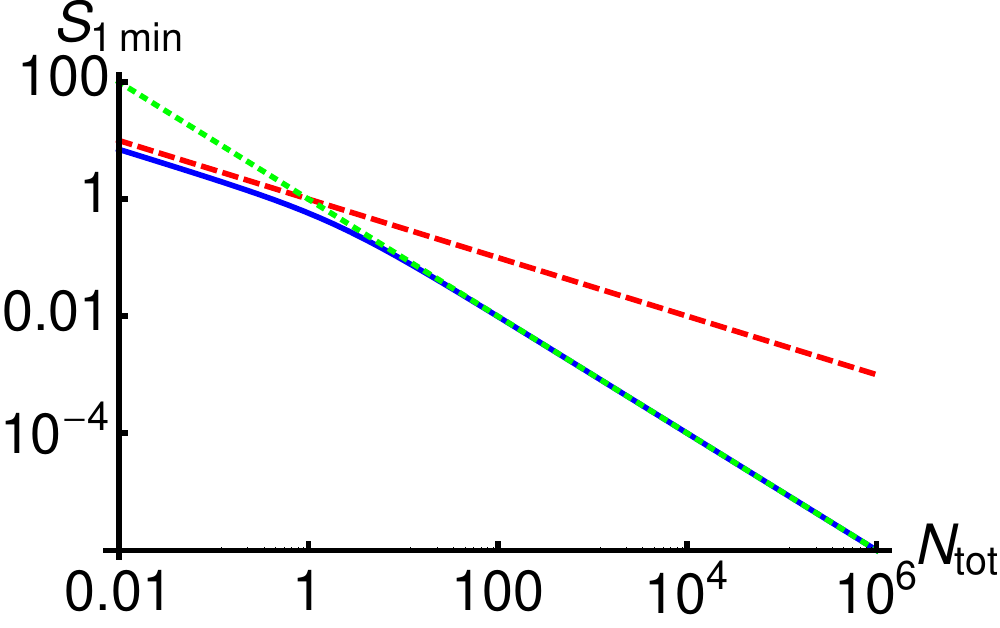}
\caption{(Color Online) The active/active interferometer with ideal
photodetection. The upper left panel shows the minimum
ideal sensitivity $S_{1\, \mathrm{min}}$ as a function
of $r_2$, for $N_{\mathrm{tot}}$ equal to $1$ (red line),
$10$ (green line), and $100$ (blue line). In the upper right panel
we show the value of $\beta$ minimizing $S_1$ as a function of $r_2$,
for the same values of $N_{\mathrm{tot}}$ of the left plot.
In the lower left panel we show the sensitivity $S_{1\, \mathrm{min}}$
as a function of $\phi$, for $N_{\mathrm{tot}}$ equal to 
$0.1$ (red line), $1$ (blue line), and $10$ (orange one).
The lower right panel shows the sensitivity $S_{1\, 
\mathrm{min}}$ as a function
of $N_{\mathrm{tot}}$. The shot noise (red dashed)
and the Heisenberg limit (green dotted) are also shown
for comparison.}
\label{f10:actact}
\end{figure}

In Sect.~\ref{s5:piam}, we have seen that the sensitivity of a passive
interferometer is minimised when the amplifier in the active measurement
stage injects a large number of photons. We study the sensitivity of
the active interferometer in the same limit, since the state
arriving at the detection stage lies in the same Gaussian sector. 
In order to validate this choice, the sensitivity $S_1$ is
numerically minimised with respect to all the parameters, except for
$N_{\mathrm{tot}}$ and $r_2$.  Results are shown
in the upper left panel of 
Fig.~\ref{f10:actact}, confirming that the sensitivity
is minimized when the OPA introduces a large number of photons
before the measurement.  Thus, we study the sensitivity in the limit of
$r_2 \rightarrow \infty$.
\par
It is interesting to notice that, when $r_2 \gg 1$, the sensitivity
$S_1$ is minimised for $\beta = 1$. This is shown in the upper right
panel of Fig.~\ref{f10:actact} where the value of $\beta$ minimizing 
the sensitivity is shown as a function of $r_2$ for different 
values of $N_{\mathrm{tot}}$. In other words, as far as 
$r_2 \gg 1$, the optimal sensitivity is obtained when all the 
photons are introduced inside the interferometer with the
first amplifier. No coherent radiation is needed and 
the input signal is just the vacuum in both modes.
Since the input state is vacuum, the coherent trade-off coefficient
$\delta$ loses its meaning, and in fact we find that the sensitivity
does not longer depend on it. Moreover, it is possible to show that
the analytic expression of $S_1$, after setting $\beta = 1$ is 
given by
\begin{align}
S_{1\, \mathrm{min}}=& \Bigg\{\Big[\csc ^2(\theta_1 -\phi ) \Big(N_{\mathrm{tot}} (N_{\mathrm{tot}}+2) \cos 2 (\theta_1 - \phi )\nonumber\\
&+4\, (N_{\mathrm{tot}}+1) \sqrt{N_{\mathrm{tot}} 
(N_{\mathrm{tot}}+2)} \cos (\theta_1 -\phi )\nonumber\\
&+3\, N_{\mathrm{tot}}(N_{\mathrm{tot}}+2)+2\Big)\Big]/
\Big[2\, N_{\mathrm{tot}}
(N_{\mathrm{tot}}+2)\Big]\Bigg\}^{\frac{1}{2}}\,.
\end{align}
The sensitivity depends only on the difference between $\theta_1$ and
$\phi$ and we can thus neglect one of them, e.g. $\theta_1$. The last
parameter to consider in order to optimize the sensitivity is the phase
$\phi$. In the lower left panel of Fig.~\ref{f10:actact}, we show
$S_{1}$ as a function of $\phi$. It is possible to see that the minimum
is achieved for values of $\phi$ close to $\pi$, though the exact value
changes with $N_{\mathrm{tot}}$.  
Eventually, the analytic form of the optimal sensitivity is 
given by 
\begin{equation}
S_{1\, \mathrm{min}}(N_{\mathrm{tot}}) = 
\frac{1}{\sqrt{N_{\mathrm{tot}} (N_{\mathrm{tot}}+2)}}
\end{equation}
which approaches the Heisenberg limit $N_{\rm tot}^{-1}$ in the
high-energy regime $N_{\mathrm{tot}} \gg 1$. The behaviour 
of the sensitivity as a function of the total number of 
photons is shown in the lower right panel of Fig.~\ref{f10:actact}.
\par
We have seen that this interferometer, like the two passive interferometers
analyzed before, has an optimized sensitivity which approaches 
the Heisenberg limit. In order to achieve the minimum sensitivity,
the input should be prepared in the vacuum state, and the whole
energy has to be provided by the first amplifier. During the measurement
stage, the second amplifier has to pump as many photons as possible
to increase minimise the sensitivity.
\subsubsection{Non unit quantum efficiency}
\begin{figure}[h!]
\includegraphics[width=0.9\columnwidth]{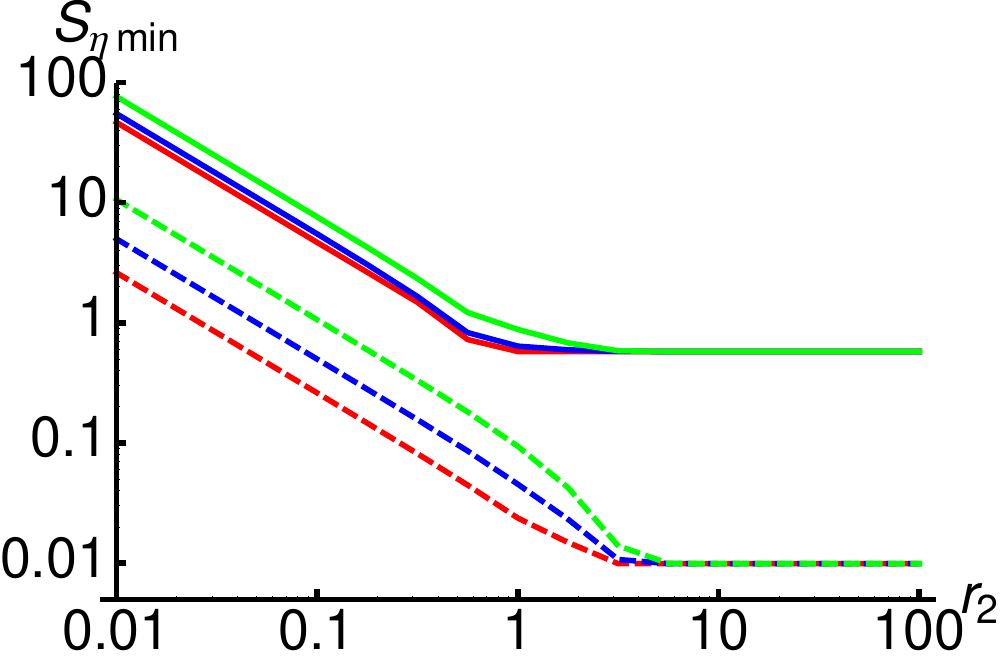}
\caption{(Color online) The active/active interferometer
with realistic photodetection. The plot shows the 
sensitivity $S_{\eta\, \mathrm{min}}$ as a function
of $r_2$, for values of $N_{\mathrm{tot}}$ equal to $1$ (solid lines),
and $100$ (dashed lines), and values of $\eta$ equal to $1$ (red lines),
$0.6$ (blue line) and $0.2$ (green ones).}
\label{CLBI_Se}
\end{figure}

When the quantum efficiency
of the detectors is lower than unit, $S_{\eta}(N_{\mathrm{tot}}, \delta,
\phi, \beta, \theta_1, r_2)$ depends on all the parameters we have
introduced before, as well as on $\eta$.
First of all, we investigate the behavior of $S_{\eta}$ as a function
of the squeezing parameter $r_2$ of the second amplifier.
To study this behavior, we perform a numerical minimization with
respect to all the parameters, except for $N_{\mathrm{tot}}$,
$\eta$ and $r_2$. In Fig.~\ref{CLBI_Se}, the minimized sensitivity
$S_{\eta\, \mathrm{min}}$ is shown as a function of $r_2$, for different
values of $\eta$ and $N_{\mathrm{tot}}$.
\par
From the plot we obtain that the sensitivity $S_{\eta}$ is minimized
when $r_2 \gg 1$, independently of the value of $N_{\mathrm{tot}}$ and
$\eta$. In particular, we find that the sensitivity is equal to
the ideal sensitivity for $r_2 \gg 1$. Therefore, if we pump into the
system a large amount of energy with the second amplifier, we balance
the losses of detectors, and are back to the ideal case.
This result is analogue to that obtained in sect.~\ref{s5:piam} for
the passive interferometer with active measurement stage.
\par
Since the sensitivity becomes equal to the ideal one for $r_2 \gg 1$,
the results of the previous section still hold. In particular, we have
Heisenberg scaling in the high-energy regime, even if the detectors 
are affected by a non-unit
quantum efficiency.
\section{Features of the active measurement stage}\label{s6:feat}
The interferometers with active measurement stage shows a particular 
feature: the sensitivity $S_{\eta}$ in the presence of non unit
quantum efficiency can be made equal to the ideal value $S_1$, 
by pumping a large number
of squeezed photons inside the system with the OPA. In the following
we further investigate this effect.
\par
We consider the active measurement scheme, 
where the detectors have the same quantum efficiency $\eta$.
The state of the modes $a$ and $b$ just before the active measurement stage is
described by a generic state $\rho_\phi$, where $\phi$ is the phase
parameter we want to estimate. Using the Heisenberg picture,
the operator $D_{+}(\eta)$ associated with
the sum photocurrent (after the OPA) can be written as (we assume, without
loss of generality, that the OPA squeezing parameter $\xi$ is real):
\begin{align}
D_{+}(\eta) = \eta \big[
(1 & + N_{\rm OPA}) \langle \Nin \rangle_{\rho_\phi}
+ N_{\rm OPA} \nonumber\\
&+\sqrt{N_{\rm OPA}(2+N_{\rm OPA})} \langle \Xab \rangle_{\rho_\phi}
\big] 
\end{align}
where $\langle \cdots \rangle_{\rho_\phi} = {\rm Tr}[\cdots\, \rho_\phi]$,
$\Nin = a^\dag a + b^\dag b$, $\Xab = a^\dag b^\dag + a b$ and
$N_{\mathrm{OPA}} = 2 \sinh^2\xi$ is the number of squeezed
photons introduced with the amplifier. It is easy to show that:
\begin{align}
\Delta D_{+}^{2}(\eta) = \eta^2 \, \Delta D_{+}^{2}
+  \eta (1-\eta) \, D_{+}
\end{align}
where $D_{+} \equiv D_{+}(1)$ and  the corresponding variance reads:
\begin{align}
\Delta D_{+}^{2} =\ & \langle D_{+}^2 \rangle_{\rho_{\phi}}
- \langle D_{+} \rangle_{\rho_{\phi}}^2 \nonumber\\
=\ &(1+N_{\rm OPA})^2 \Delta N_{\rm in}^2 +
N_{\rm OPA} (2+ N_{\rm OPA}) \Delta \Xab^2\nonumber\\
& + 2 (1 + N_{\rm OPA})\sqrt{N_{\rm OPA} (2+ N_{\rm OPA})}\nonumber\\
& \times \left(
\frac{\langle N_{\rm in} \Xab + \Xab N_{\rm in} \rangle_{\rho_\phi}}{2}-
\langle N_{\rm in}  \rangle_{\rho_\phi}
\langle \Xab \rangle_{\rho_\phi}
\right).
\end{align}
\par
The sensitivity $S_{\eta}$ for this measurement stage can be evaluated
with the Eq.\ (\ref{sens}), and it is equal to
\begin{align}\label{Se_act}
S_{\eta} =\ &\frac{\sqrt{\Delta D_+^2(\eta)}}{\left| \partial_{\phi} D_+(\eta) \right|},\\
=\ & S_1 \sqrt{1+
\frac{1-\eta}{\eta}\, \frac{D_{+}}{\Delta D_{+}^2}}\,.
\end{align}
Since $D_{+} / \Delta D_{+}^2 \propto N_{\rm OPA}^{-1}$,
it is now evident that, in the limit  $\eta N_{\mathrm{OPA}} \gg 1$, one has
$S_{\eta} \approx S_1$.
We stress that this is a completely general result, since we are not making any
assumption on the nature of the state $\rho_{\phi}$.
\section{Concluding remarks}\label{s7:con}
In this paper we have investigated the performances of optical 
interferometers involving Gaussian input signals and passive or 
active devices in the mixing stage as well as in the 
detection stage. For all the configurations we found the optimal
working point of the interferometer by optimizing over all the 
involved parameters, either characterizing the interferometer or its 
input signals.  
\par
Upon analyzing the behaviour of the QFI, we have shown that
both passive and active interferometers may achieve the Heisenberg 
scaling $\propto N_{\rm tot}^{-1}$ for a suitably optimized input
signals in the high-energy regime, with the passive ones 
performing slightly better (by a multiplicative factor). In fact, the 
passive interferometer can fully exploit the squeezing resource 
whereas, in the case of the active one, part of the squeezing is 
lost to create entanglement between the two outgoing beams leading 
to a loss of local phase sensitivity. Our results also show that 
in order to achieve the ultimate limit allowed by quantum mechanics 
one should require that one of the two beams plays the role of a 
phase reference which is indeed enhanced by the use of squeezing. 
\par
We then moved to the realistic scenario in which we have
a measurement stage based on passive or active elements and
photon number detectors, taking also into account the presence of losses
leading to a non-unit quantum efficiency. Our analytical and numerical
results have shown that in the presence of unit quantum efficiency the
symmetric configuration should be preferred: a passive (active)
interferometer should use a passive (active) detection scheme.  As one
may expect, when losses affect the detection, the Heisenberg scaling is
lost. However, we found that for both the passive and active
interferometers the presence of an OPA at the measurement stage pumping
a large number of squeezed photons allows to compensate the detrimental
effect of losses and to achieve the same sensitivity as in the ideal
casey, thus restoring again the Heisenberg
scaling.
\par
Our results show the robustness of Gaussian interferometers against 
loss, suggesting that their performances in realistic conditions may 
overcome those of the corresponding schemes involving finite
superposition of photonic states \cite{raf1,raf2}, at least
when synchronization is allowed between the sender and the
receiver.
\acknowledgments
This work has been supported by EU through the Collaborative 
Project QuProCS (Grant  Agreement 641277) and by UniMI through
the H2020 Transition Grant 15-6-3008000-625.
\appendix
\section{Sum and difference photocurrent for non unit quantum 
efficiency}\label{a1:mea}
\begin{figure}[h!]
\center
\includegraphics[width=0.75\columnwidth]{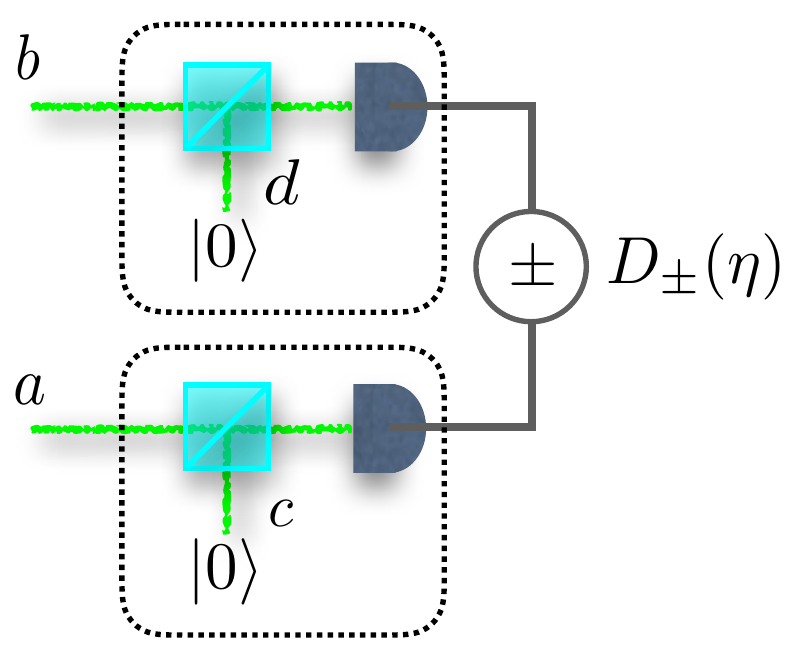}
\caption{Sum and difference photocurrents $D_+(\eta)$ 
and $D_-(\eta)$ with realistic photodetectors. Here, we use a 
beam splitter with transmission coefficient $\eta$ to model 
quantum efficiency of detectors. Since a beam splitter has two input
modes, one is filled with the radiation signal (the modes $``a"$
and $``b"$, in the scheme), while in the other (that is, the modes
$``c"$ and $``d"$) we put a vacuum state $\ket{0}$.}
\label{fa:photo}
\end{figure}

Here we address the first statistical moments of the sum- and 
the difference-photocurrent observables when the photodetectors have 
a non-unit quantum efficiency $\eta$. We model each realistic 
photodetector as an ideal one preceded by a beam splitter
of transmissivity $\eta$, with the ancillary mode placed in 
the vacuum, see Fig.\ (\ref{fa:photo}). 
\par
The first operator we consider is the sum photocurrent.
After straightforward calculation, both the mean value and
the variance of this observable can be evaluated.
The mean value is given by the equation
\begin{equation}\label{first}
\langle D_{+}(\eta) \rangle = \eta\, \langle a^{\dagger} a + b^{\dagger} b \rangle
\end{equation}
In other words, the mean value of the sum photocurrent
is equal to the total number of photons inside the signal,
scaled by the quantum efficiency of the detectors.
Instead, the variance of this observable is given by
\begin{align}
\langle \Delta D^2_{+}(\eta) \rangle =&\: \eta^2\, \Big( \langle \Delta (a^{\dagger} a)^2 \rangle + \langle \Delta (b^{\dagger} b)^2 \rangle\nonumber\\
&+ 2\; \langle a^{\dagger} a \otimes b^{\dagger} b \rangle - 2\; \langle a^{\dagger} a \rangle \langle b^{\dagger} b \rangle \Big)\nonumber\\
&+ \eta ( 1 - \eta )\, \langle a^{\dagger} a + b^{\dagger} b \rangle
\end{align}
\par
The other observable we consider is the difference photocurrent $D_-(\eta)$.
Its mean value is given by
\begin{equation}
\langle D_{-}(\eta) \rangle = \eta\, \langle a^{\dagger} a - b^{\dagger} b \rangle
\end{equation}
In this case, the mean value of $D_-(\eta)$ is the difference
between the number of photons collected by the two photodetectors.
Again, the whole expression is scaled by $\eta$.
In conclusion, the variance of the difference photocurrent can be evaluated
\begin{align}\label{last}
\langle \Delta D^2_{-}(\eta) \rangle =&\: \eta^2\, \Big( \langle \Delta (a^{\dagger} a)^2 \rangle + \langle \Delta (b^{\dagger} b)^2 \rangle\nonumber\\
&- 2\; \langle a^{\dagger} a \otimes b^{\dagger} b \rangle + 2\; \langle a^{\dagger} a \rangle \langle b^{\dagger} b \rangle \Big)\nonumber\\
&+ \eta ( 1 - \eta )\, \langle a^{\dagger} a + b^{\dagger} b \rangle
\end{align}
\par
In the equations from Eq.\ (\ref{first}) to Eq.\ (\ref{last}),
the brackets $\langle \ldots \rangle$ represent the mean value
on the state $\rho$ coming from the interferometer, that is to say,
$\langle \ldots \rangle = \tr{\rho\, \ldots}$.

\end{document}